\documentclass[preprint,showpacs,showkeys, amsmath,amssymb, aps, prfluids,floatfix,superscriptaddress]{revtex4-2}

\usepackage{color}
\usepackage{amsmath}
\usepackage{graphicx}% Include figure files
\usepackage{dcolumn}% Align table columns on decimal point
\usepackage{bm}% bold math
\usepackage{hyperref}
\usepackage{etoolbox}
\usepackage{tikz}
\usepackage{braket}

\usepackage{soul}

\def\hh{\mathfrak{h}}
\def\abs#1{\left|#1\right|}
\def\vp{{\bm{p}}}
\def\hata{{\hat{a}}}
\def\hatb{{\hat{b}}}
\def\hatc{{\hat{c}}}

\def\hatR{{\hat{R}}}
\def\hatr{{\hat{r}}}
\def\hatz{{\hat{z}}}
\def\hvarphi{{\hat{\varphi}}}
\def\htheta{{\hat{\theta}}}

\def\wi{{\widetilde{\imath}}}
\def\wj{{\widetilde{\jmath}}}
\def\wk{{\widetilde{k}}}

\def\vp{{\bf{p}}}
\def\vu{{\bf{u}}}

\def\vxi{{\bf{\xi}}}

\def\hh{\mathfrak{h}}

\def\hata{{\hat{a}}}
\def\hatb{{\hat{b}}}
\def\hatc{{\hat{c}}}

\def\hatR{{\hat{R}}}
\def\hatz{{\hat{z}}}
\def\hvarphi{{\hat{\varphi}}}

\def\wi{{\widetilde{\imath}}}
\def\wj{{\widetilde{\jmath}}}
\def\wk{{\widetilde{k}}}
\def\well{{\widetilde{\ell}}}

\def\vp{{\bm{p}}}
\def\vu{{\bm{u}}}

\def\vxi{{\bm{\xi}}}

\def\hh{\mathfrak{h}}

\begin{document}

%\preprint{}

 \title{Mesoscopic Lattice Boltzmann modeling of dense gas flows in curvilinear geometries}


\author{Sergiu Busuioc}
\email[]{sergiu.busuioc@e-uvt.ro}
\affiliation{Department of Physics, West University of Timişoara\\
Bd. Vasile P\^{a}rvan 4, 300223 Timişoara, Romania}
\affiliation{Institute for Advanced Environmental Research, West University of Timişoara\\
Bd. Vasile P\^{a}rvan 4, 300223 Timişoara, Romania}
{\large }
\date{\today}

\begin{abstract}
We derive the Enskog equation utilizing orthonormal vielbein fields, enabling the utilization of arbitrary coordinate systems to characterize spatial geometry. Additionally, we employ an adapted coordinate system in the momentum space, connected to the physical space through vielbeins. Within this framework, the momentum component perpendicular to a curved boundary can be treated as an independent one, facilitating the application of half-range Gauss-Hermite quadratures. We develop an appropriate finite-difference Lattice Boltzmann model and validate it against a DSMC-like particle-based method for solving the Enskog equation in curvilinear geometries. Our test scenarios include cylindrical Couette flow, cylindrical Fourier flow between coaxial cylinders, and spherical Fourier flow between concentric spheres. Excellent agreement between the two approaches is observed throughout the parameter range and curvature-specific effects are well captured.
\end{abstract}

\maketitle

\section{\label{sec:intro}Introduction}

Rarefied gas flows characterized by non-negligible Knudsen numbers ($Kn$), representing the ratio between the mean free path of fluid molecules and the characteristic length of the flow domain, have traditionally been investigated numerically using the Boltzmann equation, treating fluid constituents as point particles. However, when the mean free path becomes comparable to the particle size, the finite molecular size's influence becomes critical~\cite{FK72}. Practical applications involving this scenario are the single-bubble sonoluminescence~\cite{BHL02}, high-pressure shock tubes~\cite{PH01}, flows through microfabricated nanomembranes~\cite{HPWSAGNB06} and the gas extraction in unconventional reservoirs~\cite{WLRZ16,SPC17}. The Enskog equation\cite{enskog22} provides a means to extend the kinetic theory description beyond the dilute-gas Boltzmann limit. Unlike the Boltzmann approach, the Enskog equation accounts for the finite size of gas molecules, incorporating space correlations between colliding molecules, molecular mutual shielding, and reduction of available volume~\cite{enskog22, cowling70, FK72, K10, DBK2021}. Numerical solutions of the Enskog equation can be achieved through probabilistic or deterministic methodologies, mirroring the approaches used for the Boltzmann equation. Deterministic techniques, including the Monte Carlo quadrature method \cite{FS93}, the fast spectral method \cite{WZR15, WLRZ16}, and the Fokker-Planck approximation \cite{SG17, SG19}, have been applied to tackle the Enskog equation. On the other hand, probabilistic methods have emerged, inspired by the success of the Direct Simulation Monte Carlo method (DSMC) pioneered by Alexander et al. \cite{AGA95}, Montanero et al. \cite{MS96}, and Frezzotti \cite{F97b}. These methods have been applied to investigate dense gases near solid walls in micro- and nano-channels~\cite{D87,DM97,F97b,F99,NFSJMH06,SGLBZ20,WGLBZ23}. Its extension to weakly attracting hard-sphere systems has demonstrated efficacy in characterizing various phenomena, including monoatomic \cite{FGL05, KKW14, FBG19, BGLS20}, polyatomic fluids \cite{Bruno2019, BG20}, mixtures \cite{KSKFW17}, the formation and rupture of liquid menisci in nanochannels \cite{BFG15}, as well as the growth/collapse dynamics of spherical nano-droplets/bubbles \cite{BFG23}.
Despite their reliability, these methods are computationally intensive. To address this, a common approach is to simplify the Enskog collision integral by expanding it into a Taylor series. This simplification, utilized in models like Lattice Boltzmann (LB) models, has been successful in investigating non-ideal gases~\cite{L98,L00,MM08} and multiphase flows~\cite{HD02}. More recently, the simplified Enskog collision operator has been successfully implemented in various solvers, including the discrete unified gas kinetic scheme (DUGKS)~\cite{CWWC23}, the discrete velocity method (DVM)~\cite{WWHLLZ20,WGLBZ23},  the discrete Boltzmann method (DBM)~\cite{ZXQWW20, GXLLSS22}, the double-distribution Lattice Boltzmann model (DDLB)~\cite{HWA21} and the Finite-difference Lattice Boltzmann (FDLB) models~\cite{B23,BS24}. These solvers offer computationally efficient alternatives for investigating micro-scale flow phenomena while maintaining reasonable accuracy.

In this paper, we further extend the recently introduced finite difference Lattice Boltzmann (FDLB) model for bounded dense gas flows~\cite{BS24}, to tackle flows in curvilinear geometries. In the literature, it's a common approach to leverage the symmetries inherent in non-Cartesian geometries through the use of curvilinear geometry-fitted coordinates\cite{GZ03,LZ09,LXZLS14,W15,HH17,VMM19}. These coordinates can be selected such that the boundary aligns orthogonally with the unit vector along one of the curvilinear axes. To ensure the implementation of the half-range quadratures\cite{AS16a,BA19} along the direction perpendicular to the boundary, an additional step is necessary: the momentum space must be adjusted to align with the new coordinate system, ensuring that the momentum vector components consistently align with the unit vectors corresponding to the curvilinear coordinates. In Ref.~\cite{CEM13}, the relativistic Boltzmann equation has been expressed in conservative form with respect to a vielbein (i.e.~tetrad) field and a general choice for the parametrization of the momentum space. Later, in Ref.~\cite{BA19}, a formulation of the Boltzmann equation with respect to general coordinates has been presented. In order to keep the momentum space tied to the new coordinate frame, an orthonormal vielbein field (i.e. triad consisting of the non-holonomic unit vectors of the coordinate frame) has been employed, with respect to which the momentum space degrees of freedom are defined. The resulting Boltzmann equation contains inertial forces that ensure that freely streaming particles travel along straight lines in the original Cartesian geometry. Following \cite{BA19}, we apply the same vielbein framework to the Enskog equation, in order to tackle the flow of dense gases in curvilinear geometries. This framework has been successfully applied to the torus geometry involving binary fluids~\cite{ABWPK19}.

Furthermore, we demonstrate the applicability of our proposed scheme to dense gas flows enclosed inside curved boundaries by considering the cylindrical Couette flow and cylindrical Fourier flow between coaxial cylinders, as well as the spherical Fourier flow between concentric spheres. In particular, we use the cylindrical/spherical coordinates to parametrize the flow domain, such that the boundaries are orthogonal to the radial directions $R$ and $r$, for cylindrical and spherical coordinates systems, respectively. After expressing the momentum space vectors with respect to the unit vectors, we employ the mixed-quadrature lattice Boltzmann (LB) models introduced in Ref.~\cite{AS16a}. The mixed quadrature models allow the half-range or full-range Gauss-Hermite quadratures to be independently used on the coordinate axes of the momentum space. The implementation of the inertial forces requires the theory of distributions, as discussed in Ref.~\cite{ASFB17arxiv,BA19}.

The structure of the paper is as follows. For the readers's convenience, in the first part of Sec. \ref{sec:vielbeinLB}, the simplified Enskog equation is presented along with the Shakhov collision term in Cartesian coordinates. The extension to general coordinates and the vielbein framework is discussed further as well as the cylindrical and spherical cases specific to the problems addressed in this study. The Finite Difference Lattice Boltzmann (FDLB) model with mixed quadratures used to numerically solve the simplified Enskog equation in cylindrically and spherically symmetric setups is briefly introduced in Sec.~\ref{sec:FDLB}. This model relies on half-range Gauss-Hermite quadratures in order to account for the boundary-induced discontinuities. The computer simulation results for the cylindrical Couette and the Fourier flows between coaxial cylinders, as well as for the Fourier flow between concentric spheres are reported in Sec.~\ref{sec:results}. We conclude the paper in Sec.\ref{sec:conclusions}.

\section{\label{sec:vielbeinLB}Lattice Boltzmann model in curvilinear geometries}

The Enskog equation, proposed in 1922~\cite{enskog22}, describes the evolution of a system consisting of rigid spherical molecules. Unlike Boltzmann's equation, which assumes molecules as point-like particles subjected to local collisions, Enskog's equation takes into account also the volume of the fluid molecules. This volume restricts the free movement space available to each particle, leading to an increased number of collisions. Additionally, the collisions between particles are non-local, occurring when the centers of the two colliding molecules are separated by one molecular diameter. The Enskog equation can be written as follows~\cite{cowling70, K10, DBK2021}:
\begin{multline}\label{eq:enskog}
 \frac{\partial f}{\partial t}+\frac{\bm{p}}{m} \cdot\bm{\nabla}_{\bm{x}} f + \bm{F}\cdot\bm{\nabla}_{\bm{p}} f =J_E=\int \left\{ \chi\left({\bm{x}}+\frac{\sigma}{2}{\bm{k}}\right)f({\bm{x}},\bm{p^*})f({\bm{x}} + \sigma {\bm{k}},\bm{p_1^*}) \right.\\ -
    \left.   \chi\left({\bm{x}}-\frac{\sigma}{2}{\bm{k}}\right)f({\bm{x}},\bm{p})f({\bm{x}} - \sigma {\bm{k}},\bm{p_1}) \right\}\sigma^2
    ({\bm{p_r}}\cdot{\bm{k}}) d{\bm{k}}d{\bm{p_1}},
\end{multline}
where $m$ denotes the particle mass, $\bm{F}=m\bm{a}$ represents the external body force, $f\equiv f(\bm{x},\bm{p},t)$ is the single-particle distribution function, $\sigma$ represents the molecular diameter, $\bm{p_r}=\bm{p_1}-\bm{p}$ is the relative momentum, and $\bm{k}$ is the unit vector specifying the relative position of the two colliding particles. The time dependence of the distribution function is omitted for brevity. At time $t$, the distribution function $f$ provides the number of particles located within the phase space volume $d\bm{x}d\bm{p}$ centered at the point $(\bm{x},\bm{p})$. The right-hand side of the equation above is the Enskog collision operator, denoted $J_E$.

The influence of the molecular diameter $\sigma$ on the collision frequency is embedded within the contact value of the pair correlation function $\chi$. In the standard Enskog theory (SET), $\chi \equiv \chi_{\text{\tiny SET}}$ is assessed at the contact point of two colliding particles within a fluid assumed to be in uniform equilibrium \cite{K10}. An approximate yet precise expression for $\chi_{\text{\tiny SET}}$ reads:
\begin{equation}\label{eq:chi}
\chi_{\text{\tiny SET}}[n]=\frac{1}{nb}\left(\frac{P^{hs}}{n k_B T}-1\right)=\frac{1}{2}\frac{2-\eta}{(1-\eta)^3},
\end{equation}
which is derived from the equation of state (EOS) for the hard-sphere fluid proposed by Carnahan and Starling \cite{CS69}. Here, $n$ represents the particle number density, $\eta=b \rho /4$ is the reduced particle density ($b=2\pi\sigma^3/3m$), $P^{hs}$ is the pressure of the hard-sphere fluid, $k_B$ is the Boltzmann constant, and $T$ is the temperature. The square brackets in Eq.~\eqref{eq:chi} indicate functional dependence.

In the revised Enskog theory (RET), the fluid is assumed to be in a non-uniform equilibrium state \cite{K10, DBK2021, vBE73}, leading to a position-dependent particle number density. In this scenario, an effective approximation for the radial distribution function is obtained using the Fischer-Methfessel (FM) prescription \cite{FM80}. This involves substituting the actual value of the particle number density $n$ in Eq.~\eqref{eq:chi} with the average particle density $\overline{n}$ computed over a spherical volume of radius $\sigma$, centered at $\bm{x}-\frac{\sigma}{2} {\bm{k}}$:
\begin{equation}\label{eq:chi_ret}
\chi_{\mbox{\tiny RET-FM}}\left[n\Big(\bm{x}-\frac{\sigma}{2} {\bm{k}}\Big)\right]=\chi_{\mbox{\tiny SET}}\left[\overline{n}\Big(\bm{x}-\frac{\sigma}{2} {\bm{k}}\Big)\right].
\end{equation}
The average particle density $\overline{n}$ is defined as:
\begin{equation}\label{eq:fm}
\overline{n}(\bm{x})=\frac{3}{4\pi \sigma^3}\int_{\mathbb{R}^3} n(\bm{x}')w(\bm{x},\bm{x}')\,d\bm{x}',\quad
w(\bm{x},\bm{x}')=\left\{
\begin{array}{cc}
1, &\qquad \|\bm{x}'-\bm{x}\|<\sigma, \\
0, & \qquad \|\bm{x}'-\bm{x}\|\geq\sigma.
\end{array}
\right.
\end{equation}

The Enskog collision operator $J_E$ can be viewed as an extension of the Boltzmann collision operator which accounts for particles with spatial extent. As the molecular diameter $\sigma$ tends towards zero, the contact value of the pair correlation function approaches unity ($\chi\rightarrow 1$), thereby recovering the Boltzmann collision operator \cite{cowling70,K10}.

Assuming that the pair correlation function $\chi$ and the distribution functions appearing in the Enskog collision integral $J_E$ are smooth around the contact point $\bm{x}$, we can approximate these functions using a Taylor series expansion around $\bm{x}$, leading to the simplified Enskog collision operator $J_E\approx J_0+J_1$~\cite{cowling70,K10}:
\begin{eqnarray}
 J_0 \equiv J_0[f] &=& \chi\int (f^*f_1^*-ff_1)\sigma^2({\bm{p_r}}\cdot{\bm{k}}) d{\bm{k}}d{\bm{p_1}},\\
 J_1 \equiv J_1[f] &=& \chi\sigma\int\bm{k}(f^*\bm{\nabla}f_1^*+f\bm{\nabla} f_1)\sigma^2({\bm{p_r}}\cdot{\bm{k}}) d{\bm{k}}d{\bm{p_1}}\nonumber\\
 &+&\frac{\sigma}{2}\int\bm{k}\bm{\nabla}\chi(f^*f_1^* + ff_1)\sigma^2({\bm{p_r}}\cdot{\bm{k}}) d{\bm{k}}d{\bm{p_1}}.
\end{eqnarray}
The term $J_0[f]$ corresponds to the conventional collision term of the Boltzmann equation multiplied by $\chi$ and is treated as such by applying the relaxation time approximation. In this study, we utilize the Shakhov collision term~\cite{shakhov68a,shakhov68b}:
\begin{equation}\label{eq:j0}
J_0[f]=-\frac{1}{\tau}(f-f^S),
\end{equation}
where $\tau$ represents the relaxation time and $f^S$ is the equilibrium Maxwell-Boltzmann distribution multiplied by a correction factor~\cite{shakhov68a,shakhov68b,GP09,ASS20}:
\begin{equation}
f^S=f^{\text{\tiny MB}}\left[1 + \frac{1-\text{Pr}}{P_i k_B T}\left( \frac{\bm{\xi}^2}{5 m k_B T}-1 \right)\bm{\xi}\cdot \bm{q} \right].
\end{equation}
The Maxwell-Boltzmann distribution $f^{\text{\tiny MB}}$ is defined as
\begin{equation}\label{eq:fmb}
f^{\text{\tiny MB}}=\frac{n}{(2 m \pi k_B T)^{3/2}}\exp{\left(-\frac{\bm{\xi}^2}{2 m k_B T}\right)},
\end{equation}
and the heat flux $\bm{q}$ is obtained using $\bm{q}=\int d^3p f \frac{\bm{\xi}^2}{2m}\frac{\bm{\xi}}{m},$ where $\bm{\xi}=\bm{p}-m\bm{u}$ represents the peculiar momentum, $\text{Pr}=c_P\mu/\lambda$ denotes the Prandtl number, $c_P=5k_B/2m$ is the specific heat at constant pressure, $\mu$ is the shear viscosity, $\lambda$ is the thermal conductivity, and $P_i=\rho {\cal{G}} T=n k_B T$ is the ideal gas equation of state, with $\mathcal{G}$ being the specific gas constant.
It is important to note that although the Shakhov model does not guarantee non-negativity of the correction factor and the H-theorem has not been proven, the model has been successfully implemented and its accuracy has been tested through comparisons with experimental~\cite{S02,S03,GP09} or DSMC~\cite{AS18, ZXZCW19, ASS20, TWS20} results.

The term $J_1[f]$ can be approximated by replacing the distribution functions  ($f^*,f_1^*,f,f_1$) with their corresponding equilibrium distribution functions and integrating over $\bm{k}$ and $\bm{p_1}$ to obtain~\cite{cowling70,K10}:
\begin{multline}\label{eq:j1}
 J_1[f]\approx J_1[f^{\text{\tiny MB}}]=-b \rho \chi f^{\text{\tiny MB}} \left\{\bm{\xi}\cdot\left[\bm{\nabla}\ln(\rho^2 \chi T)+\frac{3}{5}\left(\zeta^2-\frac{5}{2}\right)
 \bm{\nabla}\ln T\right]\right.\\
 \left. + \frac{2}{5}\left[ 2\bm{\zeta}\bm{\zeta}\bm{:\nabla u} + \left(\zeta^2-\frac{5}{2} \right)\bm{\nabla\cdot u} \right]
 \right\},
\end{multline}
where $\bm{\zeta}=\bm{\xi}/\sqrt{2RT}$. Incorporating these approximations, the Enskog equation Eq.~\eqref{eq:enskog} can be expressed as:
\begin{equation}\label{eq:enskog_approx_cart}
\frac{\partial f}{\partial t}+\frac{\bm{p}}{m}\bm{\nabla}{\bm{x}}f + \bm{F}\cdot\bm{\nabla}_{\bm{p}} f= -\frac{1}{\tau}(f-f^S)+J_1[f^{\text{\tiny MB}}].
\end{equation}

\subsection{Enskog equation in curvilinear coordinates\label{sec:ensk_curvilinear}}

In certain situations, it is convenient to introduce a set of arbitrary coordinates $\{x^{\widetilde{1}}, x^{\widetilde{2}}, x^{\widetilde{3}}\}$, where $x^\wi \equiv x^\wi(x, y, z)$ (in this paper, we restrict our analysis to time-independent coordinate transformations). This coordinate transformation
induces a metric $g_{\wi\wj}$, as follows:
\begin{equation}
 ds^2 = \delta_{ij} dx^i dx^j = dx^2 + dy^2 + dz^2
 = g_{\wi\wj} dx^\wi dx^\wj,\label{eq:ds2}
\end{equation}
such that
\begin{equation}
 g_{\wi\wj} = \delta_{ij} \frac{\partial x^i}{\partial x^\wi} \frac{\partial x^j}{\partial x^\wj}.
\end{equation}
The Enskog equation Eq.~\eqref{eq:enskog} can be written with respect to these new coordinates as follows:
\begin{equation}
 \frac{\partial f}{\partial t} + \frac{p^\wi}{m} \frac{\partial f}{\partial x^\wi} +
 \left(F^\wi - \frac{1}{m} \Gamma^\wi{}_{\wj\wk} p^\wj p^\wk\right)
 \frac{\partial f}{\partial p^\wi} = J_E[f],\label{eq:ensk_cov}
\end{equation}
where the components $p^\wi$ and $F^\wi$ (with respect to the new coordinates) are related to the components $p^i$ and $F^i$ (expressed with respect to the old coordinates) through $ p^\wi = \frac{\partial x^{\wi}}{\partial x^i} p^i,\quad F^\wi = \frac{\partial x^{\wi}}{\partial x^i} F^i.$ The Christoffel symbols $\Gamma^\wi{}_{\wj\wk}$ appearing in Eq.~\eqref{eq:ensk_cov} are defined as:
\begin{equation}
 \Gamma^\wi{}_{\wj\wk} = \frac{\partial x^\wi}{\partial x^\ell}
 \frac{\partial^2 x^\ell}{\partial x^\wj \partial x^\wk} \nonumber
 = \frac{1}{2}g^{\wi\well} \left(\partial_\wk g_{\well\wj} + \partial_\wj g_{\well \wk} -
 \partial_\well g_{\wj\wk}\right).\label{eq:christoffel}
\end{equation}

The above formalism is sufficient to adapt the coordinate system to a curved boundary. However, the transition to an LB model is not straightforward, since the momentum space has an intrinsic dependence on the coordinates. Indeed, the Maxwellian distribution $f^{\text{\tiny MB}}$ \eqref{eq:fmb} expressed with respect to the new coordinates, reads:
\begin{equation}
 f^{\text{\tiny MB}} = \frac{n}{(2\pi m T)^{\frac{3}{2}}} \exp\left[-\frac{g_{\wi\wj} (p^\wi - mu^\wi)(p^\wj - mu^\wj)}{2mT}\right],
 \label{eq:feq_wi}
\end{equation}
while its moments are calculated as:
\begin{equation}
 M^{\wi_1, \dots \wi_n}_{\text{\tiny MB}} =
 \sqrt{g} \int d^3\widetilde{p}\, f^{\text{\tiny MB}} p^{\wi_1} \cdots p^{\wi_n},
 \label{eq:momg}
\end{equation}
where $g$ is the determinant of the metric tensor $g_{\wi\wj}$.

In order to eliminate the burden of this metric dependence in the expression for the Maxwellian,
it is convenient to introduce a triad (vielbein) with respect to which the metric is diagonal~\cite{BA19}:
\begin{equation}
 g_{\wi\wj} dx^\wi dx^\wj = \delta_{\hata\hatb} \omega^\hata \omega^\hatb,\quad  \omega^\hata = \omega^\hata_\wj dx^\wj,
\end{equation}
where $\omega^\hata$ are the triad one-forms and it follows that:
\begin{equation}
 g_{\wi\wj} = \delta_{\hata\hatb} \omega^\hata_\wi \omega^\hatb_\wj.\label{eq:gij_omega}
\end{equation}
The above equation allows three degrees of freedom for the system $\{\omega^\hata_\wj\}$,
corresponding to the invariance of the right-hand side of Eq.~\eqref{eq:gij_omega} under rotations
with respect to the hatted indices. It is possible to define triad vectors dual to the above
one-forms by introducing the following inner product~\cite{BA19}:
\begin{equation}
 \braket{\omega^\hatb, e_\hata} \equiv e_\hata^\wi \omega^\hatb_\wi = \delta^\hatb{}_\hata,\quad \text{where}\quad e_\hata = e_\hata^\wi \frac{\partial }{\partial x^\wi}.
\end{equation}
Using the above triad, the components of vectors can be expressed as $ p^\hata = \omega^\hata_\wi p^\wi,$ such that
$ g_{\wi\wj} p^\wi p^\wj = \delta_{\hata\hatb} p^\hata p^\hatb,$ resulting in the following Maxwellian:
\begin{equation}
 f^{\text{\tiny MB}} = \frac{n}{(2\pi m T)^{\frac{3}{2}}} \exp\left[-\frac{\delta_{\hata\hatb} (p^\hata - mu^\hata)(p^\hatb - mu^\hatb)}{2mT}\right],
 \label{eq:feq_triad}
\end{equation}
where the metric dependence disappeared. Its moments can be written as:
\begin{equation}
 M^{\hata_1, \dots \hata_s}_{\text{\tiny MB}} =
 \int d^3\hat{p}\, f^{\text{\tiny MB}} p^{\hata_1} \cdots p^{\hata_s}.
 \label{eq:momeq}
\end{equation}

Also, the moments of the distribution function $f$ are given by:
\begin{equation}
 M^{\hata_1, \dots \hata_s} =
 \int d^3\hat{p}\, f\, p^{\hata_1} \cdots p^{\hata_s}.
 \label{eq:mom}
\end{equation}

In curvilinear coordinates, the Enskog equation reads:
\begin{equation}
 \frac{\partial f}{\partial t} + \frac{p^\hata}{m} e_\hata^\wi \frac{\partial f}{\partial x^\wi} +
 \left(F^\hata - \frac{1}{m} \Gamma^\hata{}_{\hatb\hatc} p^\hatb p^\hatc\right)
 \frac{\partial f}{\partial p^\hata} = J_E[f],
 \label{eq:enskog_triad}
\end{equation}

Here the Enskog collision is replaced with the simplified Enskog collision operator:
\begin{multline}\label{eq:enskog_approx}
 J_E[f]=J_0[f]+J_1[f]= -\frac{1}{\tau}(f-f^S)-
 b \rho \chi f^{\text{\tiny MB}} \left\{\xi^\hata\cdot\left[\nabla_\hata\ln(\rho^2 \chi T)+\frac{3}{5}\left(\zeta^2-\frac{5}{2}\right)
\nabla_\hata\ln T\right]\right.\\
 \left. + \frac{2}{5}\left[ 2\zeta^\hata\zeta^\hata:\nabla_\hata u^\hata + \left(\zeta^2-\frac{5}{2} \right)\nabla_\hata\cdot u^\hata \right]
 \right\},
\end{multline}
where $\zeta^\hata=\xi^\hata/\sqrt{2RT}$.

The relations between the distribution function $f$ and the
particle number density $n$, macroscopic velocity $u^\hata$,
stress tensor $T^{\hata\hatb}$ and heat flux $q^\hata$ are listed below:
\begin{subequations}\label{eq:macro_def}
\begin{align}
 n =& \int d^3\hat{p} \, f,\qquad\qquad
 u^\hata = \frac{1}{\rho} \int d^3\hat{p} \, f\,p^\hata, \label{eq:macro_def_u}\\
 T^{\hata\hatb} =& \int d^3\hat{p} \, f\,\frac{\xi^\hata \xi^\hatb}{m},\qquad
 q^\hata = \int d^3\hat{p} \, f\,\frac{\vxi^2}{2m} \frac{\xi^\hata}{m},\label{eq:macro_def_q}
\end{align}
\end{subequations}
where $\rho = mn$, $\xi^\hata = p^\hata - mu^\hata$,
$\vxi^2 = \delta_{\hata\hatb} \xi^\hata \xi^\hatb$
and the internal energy is $e = \frac{1}{2n} T^\hata{}_{\hata}$.
The macroscopic fluid equations are:
\begin{subequations}\label{eq:macro}
\begin{gather}
 \frac{Dn}{Dt} + n (\nabla \cdot \vu) = 0,\qquad
 \rho \frac{Du^\hata}{Dt} = nF^{\hata} - \nabla_{\hatb} T^{\hata\hatb},\label{eq:macro_u}\\
 n\frac{De}{Dt} + \nabla_\hata q^\hata + T^{\hata\hatb} \nabla_\hata u_\hatb = 0,\label{eq:macro_T}
\end{gather}
\end{subequations}
where $D/Dt = \partial_t + u^\hata \nabla_\hata$ is the material derivative, while
$e = \frac{3}{2} T$ is the internal energy. The heat flux $q^\hata$ and the stress tensor $T^{\hata\hatb}$ are given by:
\begin{align}\label{eq:heatstress}
 q^\hata =& -\lambda \nabla^\hata T,\\
T^{\hata\hatb} =& -\mu \left(\nabla^\hata u^\hatb + \nabla^\hatb u^\hata
 - \frac{2}{3} \delta^{\hata\hatb} \nabla_\hatc u^\hatc\right),
\end{align}

The thermal conductivity $\lambda$ and the shear viscosity $\mu$ , which appear in Eqs.~({\ref{eq:heatstress}), are given by~\cite{K10}:
\begin{eqnarray}
 \mu = \tau P_i & = & \mu_0 \left[\frac{1}{\chi}+\frac{4}{5}(b\rho)+\frac{4}{25}\left(1+\frac{12}{\pi}\right)(b\rho)^2\chi\right]\, , \label{eq:viscosity} \\
 \lambda = \frac{5k_B}{2m}\frac{\tau P_i}{\text{Pr}} & = & \lambda_0 \left[\frac{1}{\chi}+\frac{6}{5}(b\rho)+\frac{9}{25}\left(1+\frac{32}{9\pi}\right)(b\rho)^2\chi\right]\, .
\end{eqnarray}
In these equations, $\mu_0=\frac{5}{16\sigma^2}\sqrt{\frac{m k_B T}{\pi}}$ and $\lambda_0=\frac{75 k_B}{64 m\sigma^2}\sqrt{\frac{m k_B T}{\pi}}$ represent the viscosity coefficient and the thermal conductivity for hard-sphere molecules at temperature $T$. For a dense gas, the Prandtl number $\text{Pr}$ is expressed as~\cite{K10}:
\begin{equation}\label{eq:prandtl}
\text{Pr}=\frac{2}{3} \,\frac{1+\frac{4}{5}b\rho\chi+\frac{4}{25}\left(1+\frac{12}{\pi}\right)(b\rho\chi)^2}{1+\frac{6}{5}b\rho\chi+\frac{9}{25}\left(1+\frac{32}{9\pi}\right)(b\rho\chi)^2}.
\end{equation}
The dilute limit $\sigma\rightarrow0$ corresponds to $\text{Pr}=2/3$.
The Chapman-Enskog expansion of Eq.~\eqref{eq:enskog_approx} provides relationships between the relaxation time $\tau$ and the transport coefficients. In this context, the relaxation time $\tau$ is expressed as:

\begin{equation}\label{eq:tau}
\tau=\frac{\mu}{P_i}.
\end{equation}

The quantity $\mu$ encompasses both kinetic and potential contributions, which account for the flow of molecules and the collisional effects on the transfer of momentum and energy in the gas~\cite{cowling70,K10}. The relaxation time approximation effectively captures the collisional transfer resulting from non-local molecular collisions. It is worth noting that the viscosity of a dense gas with a fixed reduced density $\eta$ can be adjusted by varying the molecular diameter $\sigma$ and the number density $n$.

The Knudsen number {is defined as the ratio of the mean free path and a characteristic length (in our case the distance between cylinders/spheres):
\begin{equation}
 Kn=\frac{\lambda}{L}=\frac{1}{6\sqrt{2} \eta_0 \chi(\eta_0) C},
\end{equation}
where $C=L/\sigma$ is the confinement ratio~\cite{WZR15,SGLBZ20,WGLBZ23}.

This paper primarily concentrates on benchmarking the cylindrical Couette flow, as well as the cylindrical and spherical Fourier flow.
In these cases, the steady flow either lacks bulk motion or exhibits motion perpendicular to the direction in which the gas density varies. Consequently, the bulk viscosity does not have an impact in these cases.

\subsection{Cylindrical coordinates}

Let us specialize the formalism of section~\ref{sec:ensk_curvilinear} to the case of the Couette and Fourier flow between coaxial cylinders. To describe the geometry of this flow, it is convenient to introduce cylindrical coordinates $\{x^\wi\} = \{R, \varphi, z\}$ through $x = R\cos\varphi$ and  $y = R\sin\varphi$. The line element in cylindrical coordinates and the associated triad are:
\begin{equation}
 ds^2 = dR^2 + R^2 d\varphi^2 + dz^2, \qquad e_\hatR =\partial_R, \qquad
 e_\hvarphi = R^{-1} \partial_\varphi, \qquad e_\hatz = \partial_z.
\end{equation}

The Enskog equation when the flow is homogeneous w.r.t. $\varphi$ and $z$ reads:
\begin{multline}
 \frac{\partial f}{\partial t} + \frac{2p^\hatR}{m} \frac{\partial (fR)}{\partial R^2}
 + \frac{1}{mR} \left[(p^\hvarphi)^2 \frac{\partial f}{\partial p^\hatR} -
 p^\hatR \frac{\partial (f p^\hvarphi)}{\partial p^\hvarphi}\right]=
 \frac{1}{\tau}(f-f^S)- \\b \rho \chi f^{\text{\tiny MB}} \left\{\xi^R\cdot\left[\partial_R\ln(\rho^2 \chi T)+\frac{3}{5}\left(\zeta^2-\frac{5}{2}\right)
\partial_R\ln T\right]\right.\\
 \left. + \frac{2}{5}\left[ 2\zeta^R\zeta^R:\partial_R u^R + \left(\zeta^2-\frac{5}{2} \right)\partial_R\cdot u^R \right]
 \right\}.
\end{multline}
The advection term is implemented following Ref. \cite{FK96}.

\subsection{Spherical coordinates}

To describe the spherical Fourier flow we introduce spherical coordinates $\{x^\wi\} = \{r, \theta, \varphi\}$ through $x = r \sin \theta\cos \varphi$, $y = r \sin \theta\sin \varphi$ and $z = r \cos \theta$. The line element in spherical coordinates and its associated triad are:
\begin{multline}
ds^2 = dr^2 + r^2 (d\theta^2 + \sin^2\theta d\varphi^2),\qquad
 e_{\hatr} = \partial_r,\qquad e_{\htheta} = \frac{\partial_\theta}{r},\qquad
e_{\hvarphi} = \frac{r^{-1}}{\sin\theta} \partial_\varphi.
\end{multline}
 Assuming that the flow is homogeneous w.r.t. $\theta$ and $\varphi$, the Boltzmann equation
reads:
\begin{multline}
 \frac{\partial f}{\partial t} + \frac{3p^\hatr}{m} \frac{\partial (fr^2)}{\partial r^3}
 + \frac{1}{mr} \left[\left((p^\htheta)^2 +(p^\hvarphi)^2\right) \frac{\partial f}{\partial p^\hatr} -
 p^\hatr\left(\frac{\partial (f p^\htheta)}{\partial p^\htheta} + \frac{\partial (f p^\hvarphi)}{\partial p^\hvarphi}\right)\right]=
  \frac{1}{\tau}(f-f^S)- \\b \rho \chi f^{\text{\tiny MB}} \left\{\xi^r\cdot\left[\partial_r\ln(\rho^2 \chi T)+\frac{3}{5}\left(\zeta^2-\frac{5}{2}\right)
\partial_r\ln T\right]\right.\\
 \left. + \frac{2}{5}\left[ 2\zeta^r\zeta^r:\partial_r u^r + \left(\zeta^2-\frac{5}{2} \right)\partial_r\cdot u^r \right]
 \right\}.
\end{multline}
 The advection term is implemented following Ref. \cite{DDK02}.

\section{Mixed quadrature LB models}\label{sec:FDLB}

In this paper we employ the mixed quadrature LB models, introduced in Refs.~\cite{AS16a,AS16b,ASFB17arxiv,BA19}, to obtain the numerical solution of
the Enskog equation in cylindrically and spherically symmetric flows using the simplified Enskog collision operator.
We briefly describe these models and encourage the reader to refer to the above-mentioned references for more details.
In the case of the cylindrical Couette flow, the dynamics along the $z$ direction is straightforward. In this context, it is advantageous to integrate out the trivial degrees of freedom in the momentum space at the level of the model equation. The details can be consulted in Ref. \cite{BA19}. As discussed in Ref.~\cite{SYC06}, in our Gauss-Hermite quadrature-based model the moments of the distribution function are recovered
up to a certain expansion order $N$. As such, the resulting evolution of moments up to order $N$ of $f$ is equivalent to that obtained through a Grad’s expansion procedure. In all Gauss-Hermite quadrature-
based models, the set of distribution functions corresponding to the elements of the discrete velocity set is used instead of the moment integrals, resulting in a purely kinetic description of fluid systems.

The numerical schemes employed in this paper to solve the evolution equations are the third-order total variation diminishing (TVD) Runge-Kutta method for time-stepping\cite{SO88}, the fifth-order WENO-5 advection scheme\cite{GXZL11,JS96}, and the $4$th order central difference scheme used for gradient evaluation\cite{F88}.

Another ingredient needed is the evaluation of the average density $\bar{n}$ in Eq. \eqref{eq:fm}. In spherical coordinates one has to evaluate intersections of spherical shells in order to compute the integral in Eq.~\eqref{eq:fm}, for which we have an analytic solution. For cylindrical coordinates, one has to rely on numerically computing the intersection between a cylinder shell and a spherical shell. The procedure is detailed in Appendix \ref{appendix:volintersect}.

\subsection{Quadrature choice}

Depending on the flow regime under consideration, we will consider a mixture of full-range Gauss-Hermite (HLB) and half-range Gauss-Hermite quadratures (HHLB).
Given the choice of the quadrature (full-range or half-range Gauss-Hermite) the total number of quadrature points on the axis $a$ is $\mathcal{Q}_a = Q_a$
or $\mathcal{Q}_a = 2Q_a$, respectively, where $Q_a$ is the quadrature order.
In the non-homogeneous direction (i.e. the radial one) the half-range quadratures are employed and their orders are $Q_R$ and $Q_r$, in the cylindrical and spherical coordinate systems, respectively. In the homogeneous direction, the full-range quadratures are employed and their orders are $Q_\varphi$ and $(Q_\theta,Q_\varphi)$.
% Consider the case when the half-range Gauss-Hermite quadrature is employed in the non-homogeneous direction (i.e. the radial one), namely $\{Q_R;Q_r\}$ and full-range Gauss-Hermite quadrature for the homogeneous directions, $\{Q_\varphi;(Q_\theta,Q_\varphi)\}$, in cylindrical and spherical geometry, respectively.
We denote such a model using \cite{AS16a,ASFB17arxiv,BA19}:
\begin{align}
\nonumber \text{Cylindrical}~:&~ {\rm HHLB}(N_R; Q_R) \times {\rm HLB(N_\varphi; Q_\varphi)},\\
\nonumber \text{Spherical}~:&~   {\rm HHLB}(N_r; Q_r) \times {\rm HLB(N_\theta; Q_\theta)} \times {\rm HLB}(N_\varphi; Q_\varphi),
\end{align}
where $N_a$ represents the order of the expansion of the equilibrium distribution $f^{\text{\tiny MB}}$ with
respect to axis $a$, and $Q_a$ is the quadrature order.

The choice of the quadrature controls the momentum space integration as well as the discretization of the momentum space. In particular, the moments in Eq. \eqref{eq:mom} are evaluated as:
\begin{align}
 \text{Cylindrical}~:&~ M^{\hata_1, \dots \hata_s} = \sum_{i = 1}^{ \mathcal{Q}_R} \sum_{j = 1}^{\mathcal{Q}_\varphi } f_{ij} \prod_{\ell = 1}^s p^{\hata_\ell}_{ij}.\\
 \text{Spherical}~:&~ M^{\hata_1, \dots \hata_s} = \sum_{i = 1}^{ \mathcal{Q}_r} \sum_{j = 1}^{ \mathcal{Q}_\theta}
 \sum_{k = 1}^{ \mathcal{Q}_\varphi} f_{ijk} \prod_{\ell = 1}^s p^{\hata_\ell}_{ijk}.
 \label{eq:mom_quad}
\end{align}
A similar prescription holds for the macroscopic quantities appearing in Eq. \ref{eq:macro_def}.
In the cylindrical geometry, the discrete momenta components of $\vp_{ij} = \{p^{\hatR}_i, p^{\hvarphi}_j\}$ are
indexed on each direction separately, where $1 \le i \le \mathcal{Q}_R$ and $1 \le j \le \mathcal{Q}_\varphi$. In the spherical case, the discrete momenta components of $\vp_{ijk} = \{p^{\hatr}_i, p^{\htheta}_j, p^{\hvarphi}_k\}$ are
indexed as follows: $1 \le i \le \mathcal{Q}_r$,
$1 \le j \le  \mathcal{Q}_\theta$ and $1 \le k \le  \mathcal{Q}_\varphi$. Their components are the roots of the half-range Hermite polynomial $\hh_{Q_a}(x)$ of order $Q_a$  or of the full-range Hermite polynomials $H_{Q_a}(x)$ of order $Q_a$. In the case of the half-range quadrature, we use the convention that the indices $i$ of the quadrature points lying on the positive semi-axis have the values $1 \leq i \leq Q_a$, while the corresponding index of the points on the negative semi-axis have the values  $Q_a + 1 \leq i \leq 2 Q_a$~\cite{BA19}.

As an example, for the cylindrical coordinate system, the link between the Boltzmann distribution function $f(p^\hatR, p^\htheta, p^\hvarphi)$ and the discrete distributions $f_{ijk}$ is given through:
% \begin{equation}
%  f_{ijk} = \frac{w_i^{\hh}(Q_R) w_j^H(Q_\varphi) }
%  {\omega(p^\hatR_i) \omega(p^\hvarphi_j) }
%  f(p^\hatR_i, p^\hvarphi_j),
% \end{equation}
\begin{equation}
 f_{ijk} = \frac{w_i^{\hh}(Q_R) w_j^H(Q_\theta) w_k^H(Q_\varphi)}
 {\omega(p^\hatR_i) \omega(p^\htheta_k) \omega(p^\hvarphi_j) }
 f(p^\hatR_i, p^\htheta_j, p^\hvarphi_k),
\end{equation}
where $\omega(x) = \frac{1}{\sqrt{2\pi}} e^{-x^2/2}$ is the weight function for the half-range and full-range Hermite polynomials.
The quadrature weights for the full-range Gauss-Hermite and half-range Gauss-Hermite polynomials, denoted $w_i^H(Q_a)$ and $w_i^\hh(Q_a)$, respectively, are given by:
\begin{equation}
 w_i^H(Q_a) = \frac{Q_a!}{H_{Q_a + 1}^2(p^\hata_i)}, \qquad  w_i^\hh(Q_a) = \frac{p^\hata_i a_{Q_a - 1}^2}{\hh_{Q_a - 1}^2(p^\hata_i) [p^\hata_i + \hh_{Q_a}^2(0) / \sqrt{2\pi}]},
\end{equation}
where $ a_\ell = \frac{\hh_{\ell+1,\ell+1}}{\hh_{\ell,\ell}}$ is written in terms of the coefficients $\hh_{\ell,s}$ of $x^s$
in the polynomial expansion of $\hh_\ell(x)$\cite{BA19}.

\subsection{Force terms}

The terms involving the derivative of $f$ with respect to momentum vector components require an appropriate treatment,
since the discretization of the momentum space removes the functional dependence of $f$ on the momentum.
Based on the discussion in Ref.~\cite{ASFB17arxiv,BA19}, we write the two types of force terms as:
\begin{align}
 \left(\frac{\partial f}{\partial p^\hata}\right)_{ijk} =&
 \sum_{i' = 1}^{\mathcal{Q}_a} \mathcal{K}^a_{i,i'} f_{i'jk}, \label{eq:force1}\\
 \left(\frac{\partial (f p^\hata)}{\partial p^\hata}\right)_{ijk} =&
 \sum_{i' = 1}^{\mathcal{Q}_a} \mathcal{K}^a_{i,i'} f_{i'jk}.\label{eq:force2}
\end{align}

The matrix $\mathcal{K}^a_{i,i'}$ in Eq.\eqref{eq:force1} has the following form:
\begin{align}
\mathcal{K}^{a,H}_{i,i'} {}&= -w_i^H \sum_{\ell = 0}^{Q_a - 1} \frac{1}{\ell!}
 H_{\ell + 1}(p^{\hata}_i) H_\ell(p^{\hata}_{i'})\\
\nonumber \mathcal{K}^{a,\hh}_{i,i'} {}&=  w_i^\hh \sigma_i^\hata \Bigg\{
 \frac{1 + \sigma_i^\hata \sigma_{i'}^\hata}{2} \sum_{\ell = 0}^{Q_a - 2} \hh_\ell(\abs{p_{i'}^\hata})
\times [ \frac{\hh_{\ell,0}}{\sqrt{2\pi}} \sum_{s = \ell + 1}^{Q_a - 1} \hh_{s,0} \hh_s(\abs{p_i^\hata})\\
  &\qquad\qquad\qquad - \frac{\hh_{\ell, \ell}}{\hh_{\ell+1,\ell+1}} \hh_{\ell + 1}(\abs{p_i^\hatR})]- \frac{1}{2\sqrt{2\pi}} \Phi^{Q_R}_0(\abs{p_i^\hatR}) \Phi^{Q_R}_0(\abs{p_{i'}^\hatR})\Bigg\}.
 \end{align}
for the full-range and half-range quadrature, respectively. In the above, $\sigma_i^\hatR$ and $\sigma_{i'}^\hatR$,  are the signs of $p_i^\hatR$ and
$p_{i'}^\hatR$, respectively.
The function $\Phi^n_s(x)$ is given by $\Phi^n_s(x) = \sum_{\ell = s}^{n} \hh_{\ell,s} \hh_\ell(x)$, as defined in Ref.~\cite{ASFB17arxiv,BA19}.

The matrix $\mathcal{K}^a_{i,i'}$ involved in Eq.\eqref{eq:force2} is written as:
\begin{align}
\nonumber \mathcal{K}^{a,H}_{i,i'} =& -w_i^H \sum_{\ell = 0}^{Q_a - 2} \frac{1}{\ell!}
 H_{\ell + 1}(p_i^\hata) [H_{\ell+1}(p_{i'}^\hata) + \ell H_{\ell - 1}(p_{i'}^\hata)],\\
\nonumber \mathcal{K}^{a,\hh}_{i,i'} =& -w_i^\hh \frac{1 + \sigma_i^\hata \sigma_{i'}^\hata}{2}\sum_{\ell = 0}^{Q_a - 1} \hh_\ell(p_i^\hata)
 \Bigg[\ell\,\hh_\ell(p^\hata_{i'})+ \frac{\hh_{\ell,0}^2 + \hh_{\ell-1,0}^2}{a_{\ell-1}\sqrt{2\pi}}
 \hh_{\ell-1}(p^\hata_{j'}) \\
\nonumber  &+ \frac{1}{a_{\ell-1} a_{\ell-2}} \hh_{\ell-2}(p^\hata_{i'})\Bigg].
\end{align}
for the full-range and half-range quadrature, respectively.

\subsection{Equilibrium distribution function}\label{sec:LB:feq}

The construction of the equilibrium distribution function \eqref{eq:feq_triad} follows the prescription developed in Refs.~\cite{AS16a,AS16b,BA19}. $f^{\text{\tiny MB}}$ is factorised with respect to the momentum
space, such that:
\begin{gather}
 f^{\text{\tiny MB}} = n \,g_R(p^\hatR) g_\varphi(p^\hvarphi) g_z(p^\hatz), \nonumber\\
 g_a (p^\hata) = \frac{1}{\sqrt{2\pi m T}} \exp\left[-\frac{(p^\hata - mu^\hata)^2}{2mT}\right].
\end{gather}
Following the discretisation of the momentum space, $f^{\text{\tiny MB}}$ is replaced by
$f^{\text{\tiny MB}}_{ijk} = n\, g_{R,i} g_{\varphi,j} g_{z,k}$.
The function $g_{a,k}$ becomes \cite{AS16a,AS16b,BA19}:
\begin{align}
 g_{a,k}^H &= w_k^H \sum_{\ell = 0}^{N_a} H_\ell(p^\hata_k)
 \sum_{s = 0}^{\lfloor \ell / 2\rfloor} \frac{(mT - 1)^s (mu^\hata)^{\ell - 2s}}{2^s s! (\ell -2s)!},
 \label{eq:eq_gak_H}\\
 g_{a,k}^\hh &= \frac{w_k^\hh}{2} \sum_{s = 0}^{N_a} \left(\frac{mT}{2}\right)^{s/2}
 \Phi_s^{N_a}(\abs{p^\hata_k})
 \times \left[(1 + erf\zeta^\hata) P_s^+(\zeta^\hata) +
 \frac{2}{\sqrt{\pi}} e^{-\zeta_{\hata}^2} P_s^*(\zeta^\hata)\right],
 \label{eq:eq_gak_hh}
\end{align}
for the case of full-range (upperscript $H$) and half-range (upperscript $\hh$) quadrature, respectively.
In the above the expansion order $N_a$ is a free parameter satisfying $0 \le N_a < Q_a$.
The exact recovery of the moments \eqref{eq:momeq} for polynomials in $p^\hata$ of order less than or
equal to $N_a$ in ensured by an expansion of $g_{a,k}$ up to order $N_a$.
The expression for $g_{a,k}^\hh$ holds the following notations:
$\zeta^\hata = u^\hata \sqrt{m / 2T}$ when $p^\hata_k > 0$ and
$\zeta^\hata = -u^\hata \sqrt{m / 2T}$ when $p^\hata_k < 0$, while the polynomials $P_s^+(x)$
and $P_s^*(x)$ are defined as
\begin{gather}
 P_s^\pm(x) = e^{\mp x^2} \frac{d^s}{dx^s} e^{\pm x^2}, \nonumber\\
 P_s^*(x) = \sum_{j = 0}^{s -1} \binom{s}{j} P_j^+(x) P_{s-j-1}^-(x).
\end{gather}

\section{Numerical Results\label{sec:results}}

\begin{figure*}
\begin{tabular}{ccc}
\begin{minipage}{0.33\linewidth}
\begin{tikzpicture}
\filldraw[fill=black, draw=black] (50pt,50pt) circle (70pt);
\filldraw[fill=blue!10!white, draw=black] (50pt,50pt) circle (67.5pt);
\filldraw[fill=black, draw=black] (50pt,50pt) circle (32pt);
\filldraw[fill=blue!00!white, draw=black] (50pt,50pt) circle (30pt);

\node at (65pt,57pt) {$\mathcal{R}_{L}$};
\node at (40pt,100pt) {$\mathcal{R}_{R}$};
\draw[thick,->] (50pt,50pt) -- (50pt,117pt);
\draw[thick,->] (50pt,50pt) -- (80pt,50pt);

\node at (10pt,50pt) {$T_{w}$};
\node at (-27pt,50pt) {$T_{w}$};

\draw[->,line width=0.5mm] (85pt,65pt) arc (30:60:50pt) node at (85pt,85pt) {$U_w$};
\end{tikzpicture}
\end{minipage}
&
\begin{minipage}{0.33\linewidth}
\begin{tikzpicture}
\filldraw[fill=blue, draw=black] (0,0) circle (2.5);
\filldraw[fill=blue!10!white, draw=black] (0,0) circle (2.4);
\filldraw[fill=red, draw=black] (0,0) circle (1);

\draw[stealth-,red,thick] (0.0,1.0) -- ++(90:0.5) ;
\node[red] at (0.0,1.7) {Hot cylinder};
\draw[stealth-,blue!90,thick] (0,-2.4) -- ++(270:-0.6) ;
\node[blue] at (0.0,-1.5){Cold cylinder};
    \end{tikzpicture}
\end{minipage}
&
\begin{minipage}{0.33\linewidth}
\begin{tikzpicture}
    \begin{scope}
        \clip (0:2.47) arc (0:90:2.47) to[out=225,in=100,looseness=1.2] (-1.1,-1.1) to[out=-10,in=225,looseness=1.2] (0:2.47);
        \shade[ball color=blue!80!gray!80!black,shading angle=180] (0,0) circle (2.5);
    \end{scope}
    \shade[ball color=red!70!gray] (0,0) circle (1);
    \begin{scope}

    \end{scope}
    \begin{scope}
        \clip (0:2.4) arc (0:90:2.4) to[out=225,in=100,looseness=1.2] (-1.1,-1.1) to[out=-10,in=225,looseness=1.2] (0:2.4) -- (3,0) -- (3,-3) -- (-3,-3) -- (-3,3) -- (3,3) -- (3,0);
        \shade[ball color=blue!70!gray] (0,0) circle (2.5);
    \end{scope}
    \draw[stealth-,red,thick] (0,0) -- ++(70:2.7) node[right] {Hot sphere};
    \draw[stealth-,blue!90] (225:2) -- ++(225:1) node[below] {Cold sphere};

\end{tikzpicture}
\end{minipage}\\
(a)&(b)&(c)
\end{tabular}
\caption{Setups of the studied problems: (a) Cylindrical Couette flow, (b) Cylindrical Fourier flow and (c) Spherical Fourier flow.\label{fig:setups}}
\end{figure*}
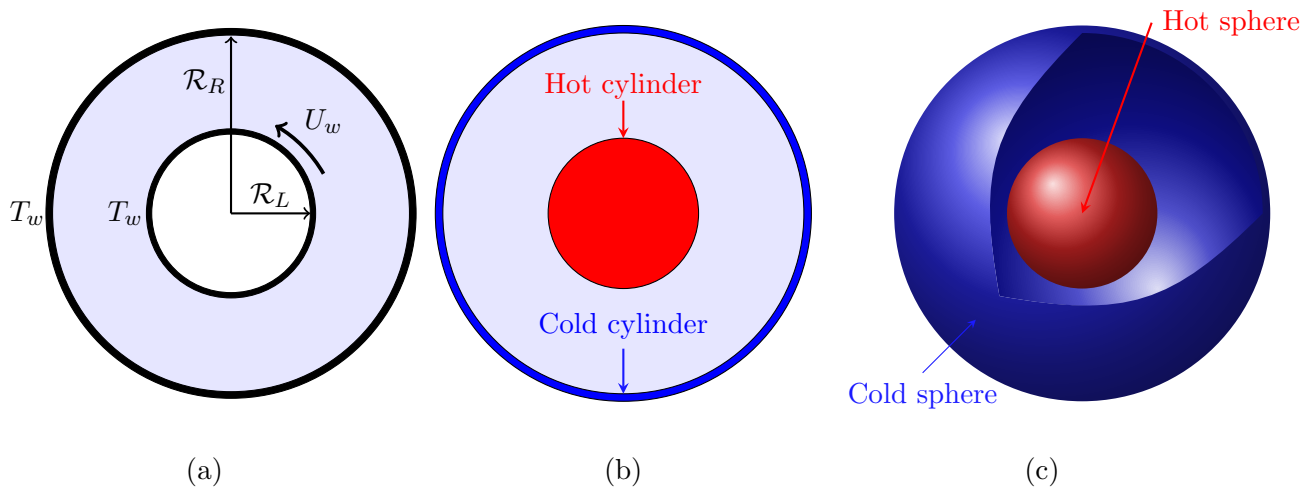

\begin{figure}

\begin{tikzpicture}
\draw[black,thick,->] (-1,1) to (-0.4,-0.2);
\node at (-0.8,-0.1) {$\mathcal{R}_{L}$};

\draw[black,thick,->] (-1,1) to (0.,2.5);
\node at (-0.9,1.8) {$\mathcal{R}_{L}^c$};

\draw[black,thick,->] (-1,1) to (2.1,-1.5);
\node at (1.,-1.1) {$\mathcal{R}_{R}$};

\draw[black,thick,->] (-1,1) to (1.9,3.);
\node at (1.,2.8) {$\mathcal{R}_{R}^c$};

\draw[black,thick,<->] (0.475,1.425) to (2.375,1.975);
\node at (1.5,2.) {$\mathcal{L}^c$};

\draw[black,thick,<->] (0,1) to (3,1.);
\node at (1.5,1.25) {$\mathcal{L}$};

\draw[black,thick] (2.325,0) circle (0.5);
\draw[black,thick,<->] (1.825,0) to (2.825,0);
\node at (2.2,0.15) {$\sigma$};

\draw[black,thick] (-0.16,-0.7) circle (0.5);
\draw[black,thick,<->] (-0.68,-0.7) to (0.33,-0.7);
\node at (-0.3,-0.55) {$\sigma$};

\draw[black,thick] (0,1) arc (0:56:2);
\draw[black,thick] (0,1) arc (0:-56:2);

\draw[black,thick,dashed] (0.5,1) arc (0:52:2.5);
\draw[black,thick,dashed] (0.5,1) arc (0:-52:2.5);

\draw[black,thick] (3,1) arc (0:50:4);
\draw[black,thick] (3,1) arc (0:-50:4);

\draw[black,thick,dashed] (2.5,1) arc (0:51:3.5);
\draw[black,thick,dashed] (2.5,1) arc (0:-51:3.5);

\end{tikzpicture}
\caption{Radial coordinates used in the paper and the particle diameter $\sigma$ for context. The dashed lines represent the computational domain, at a distance of $\sigma/2$ from the physical domain.\label{fig:coords}}
\end{figure}
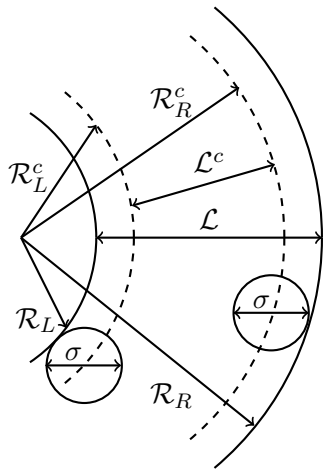

We perform tests on three distinct setups, the cylindrical Couette and Fourier flow between coaxial cylinders, and the Fourier flow between concentric spheres. The setups are presented in Fig. \ref{fig:setups}. The inner and outer cylinders/spheres radii are $\mathcal{R}_L\in\{R_L,r_L\}$ and $\mathcal{R}_R\in\{R_R,r_R\}$, respectively, and the diffuse reflection boundary conditions are applied at $\mathcal{R}_L^c=\mathcal{R}_L+\sigma/2$ and $\mathcal{R}_R^c=\mathcal{R}_R-\sigma/2$, which are the limits of the computational domain indicated by the superscript $c$. The confinement ratio is defined as $C=(\mathcal{R}_R-\mathcal{R}_L)/\sigma=\mathcal{L}/\sigma$, where $\mathcal{L}$ is the physical domain width and $\mathcal{L}^c=\mathcal{L}-\sigma$ is the width of the computational domain. These definitions are depicted graphically in Fig. \ref{fig:coords}.

The FDLB simulation results are validated using a DSMC-like particle method that extends the original Direct Simulation Monte-Carlo (DSMC) method to handle the nonlocal nature of the Enskog collision integral~\cite{F97b}. For a comprehensive explanation of the numerical scheme and an analysis of its computational complexity, please refer to Ref.~\cite{FBG19}. The DSMC framework used to solve the Boltzmann equation is maintained, with modifications made to the collision algorithm to accommodate the nonlocal structure of the Enskog collision operator. Following \cite{BFG23}, the particle stream is implemented as described in \cite{B13} and the collision process takes into account the symmetries of the flow. In the present paper, the use of particle variable weights it's not mandatory as the computational domain is relatively small, and the numerical scheme associated with variable weights conserves mass, momentum, and energy only in a statistical sense. Using constant weight, the total number of collisions is evaluated in the same manner as in Cartesian coordinates, further simplifying the process. Once a particle is selected, a vector $\hat{k}$ is drawn uniformly from the unit sphere, and the collision partner is chosen at random from the cell corresponding to the radial cell pointed by it. In general, this unit vector identifies a point off the radial axis (z-axis) and, therefore, the velocity components of the selected particle must be rotated to bring its z-component to be on the radial axis going through that point. The rotations are performed using the rotation matrix associated with the Euler angles. Please refer to \cite{BFG23} for more details.  If not stated otherwise, the time step was set to $\Delta t=10^{-3}$ and the lattice spacing (cell length for particle method) at $\Delta r=\sigma/10$. A number of $1000$ particles per cell was used in the particle method to obtain smooth profiles of macroscopic quantities.

The simulations were conducted for three values of the mean reduced density, $\eta_0\in\{0.01,0.1,0.2\}$, three values of the inner cylinder radius $\mathcal{R}_L\in\{\sigma,3\sigma,5\sigma\}$ (corresponding to the computational values $\mathcal{R}_L^c\in\{1.5\sigma,3.5\sigma,5.5\sigma\}$), while the outer cylinder radius is $\mathcal{R}_R=\mathcal{R}_L+C$ (corresponding to the computational value $\mathcal{R}_R^c=\mathcal{R}_L^c+C-\sigma$). In our simulations, we considered three values of the confinement ratio, namely $C\in\{4,7,10\}$, corresponding to $\mathcal{L}_c\in\{3\sigma,6\sigma,9\sigma\}$. The values of the associated Knudsen numbers are listed in Table \ref{tab:quad}. The objective is to specifically emphasize the unique characteristics of fluid flow when dense gas effects and confinement are involved. We fix the value of the molecular diameter at $\sigma=1$ and, as such, in the following, we will drop $\sigma$ when referring to $\mathcal{R}_L$ or $\mathcal{L}_c$. If not stated otherwise the macroscopic quantities will be plotted with respect to the reduced radius $(\mathcal{R}-\mathcal{R}_L^c)/(\mathcal{R}_R^c-\mathcal{R}_L^c)$.

Please refer to Appendix \ref{app:stationary} for a comprehensive study of gas at rest between coaxial cylinders and concentric spheres.

\begin{table*}
	\begin{center}
	\begin{ruledtabular}
%		\begin{tabular}{2\columnwidth}{@{\extracolsep{\stretch{1}}}*{7}{l||ccc|ccc|ccc}@{}}
		\begin{tabular}{l|ccc|ccc|ccc}
% 		\toprule
    $R $  &  & 4 &  &  & 7 & & & 10 &  \\ \hline
    $\eta_0$	& 0.01 & 0.1 & 0.2 & 0.01 & 0.1 & 0.2 & 0.01 & 0.1 & 0.2  \\ \hline
	$Kn$		&  2.8731 & 0.2261 & 0.0838 & 1.64179 & 0.1292 & 0.04789 & 1.1493 &  0.0904 & 0.0335
		\end{tabular}
		\end{ruledtabular}
	\end{center}
	\caption{ The Knudsen number $Kn$ for the parameters used in this study. }
	\label{tab:quad}
\end{table*}

\subsection{Cylindrical Couette flow}

\begin{figure*}
\begin{tabular}{ccc}
\includegraphics[width=0.325\linewidth]{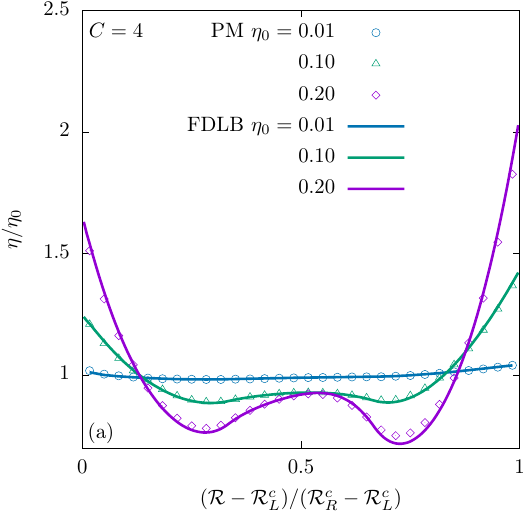}&
\includegraphics[width=0.325\linewidth]{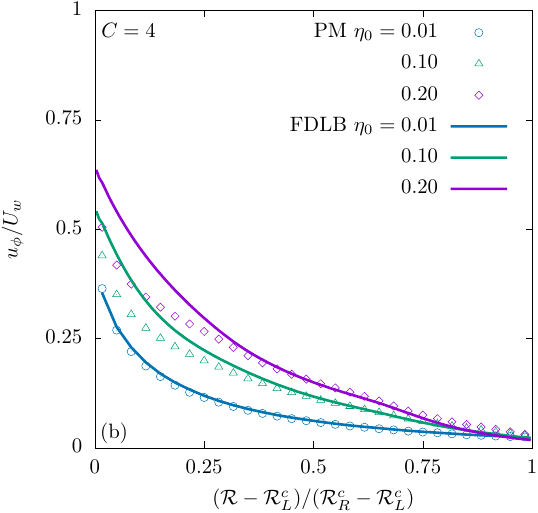}&
\includegraphics[width=0.325\linewidth]{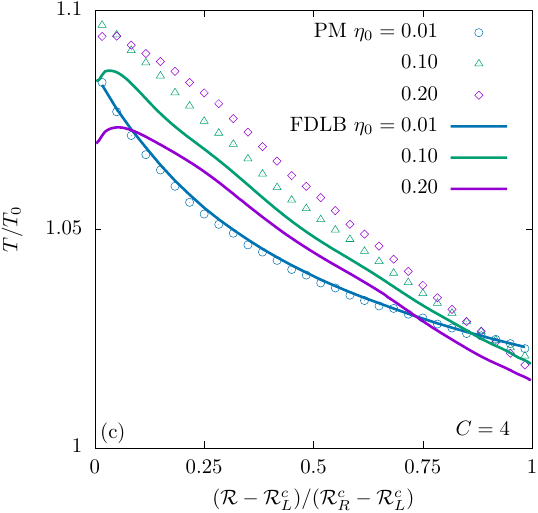}\\
\includegraphics[width=0.325\linewidth]{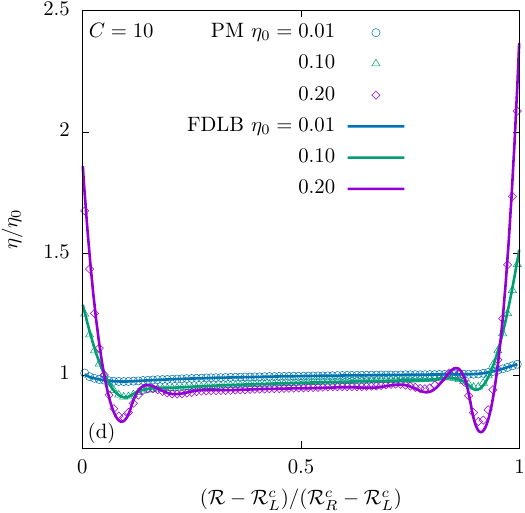}&
\includegraphics[width=0.325\linewidth]{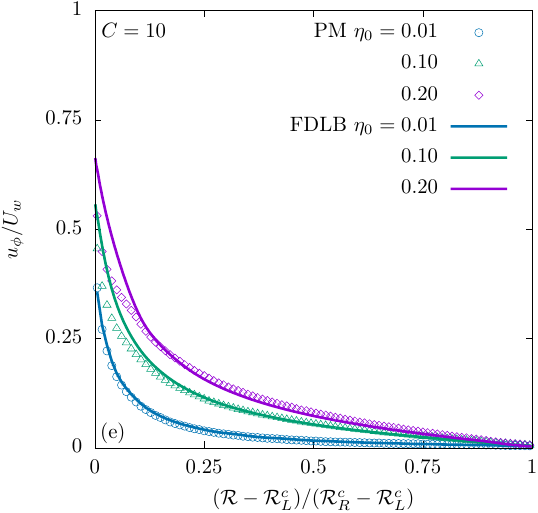}&
\includegraphics[width=0.325\linewidth]{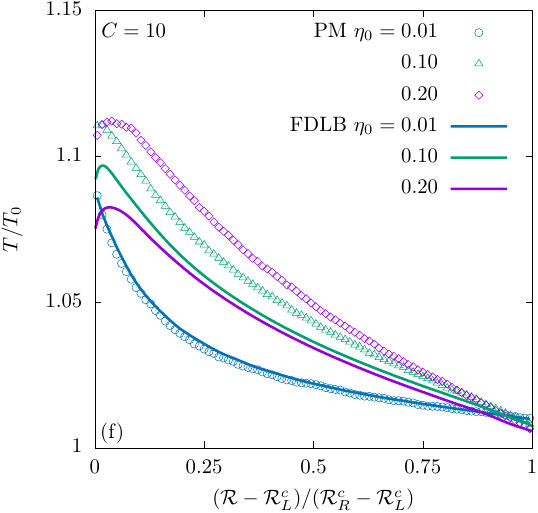}
\end{tabular}
\caption{Cylindrical Couette flow: Normalized (a,d) reduced density $\eta/\eta_0$, (b,e) azimuthal velocity $u_\phi/U_w$ and (c,f) temperature $T/T_0$, when the inner cylinder radius is $\mathcal{R}_L=1$, the confinement ratio is $C\in\{4,10\}$, for three values of the mean reduced density $\eta_0\in\{0.01,0.1,0.2\}$. \label{fig:couettecyl_rl1_vareta}}
\end{figure*}

\begin{figure*}
\begin{tabular}{ccc}
\includegraphics[width=0.325\linewidth]{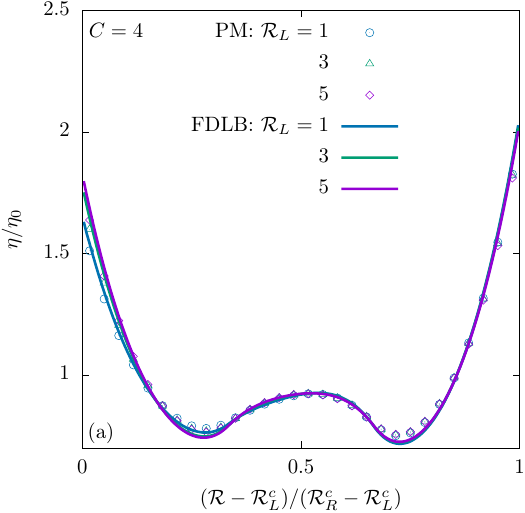}&
\includegraphics[width=0.325\linewidth]{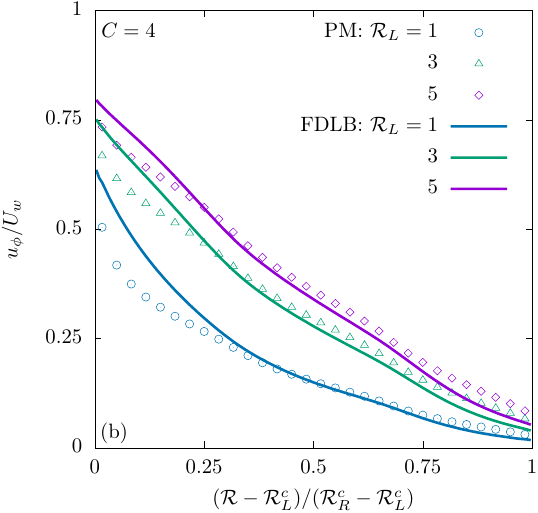}&
\includegraphics[width=0.325\linewidth]{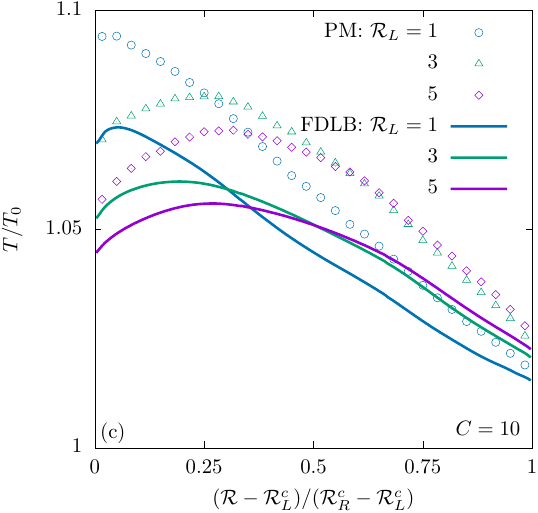}\\
\includegraphics[width=0.325\linewidth]{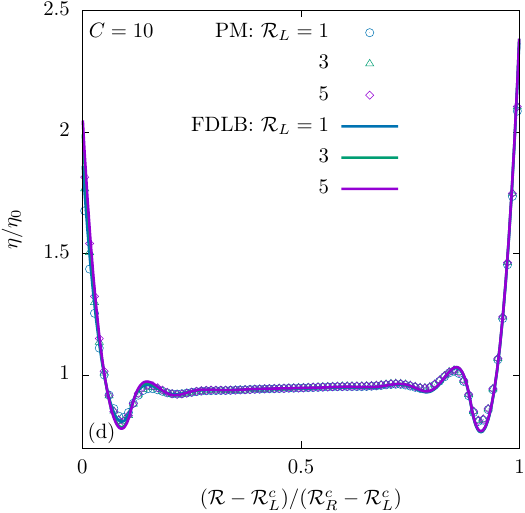}&
\includegraphics[width=0.325\linewidth]{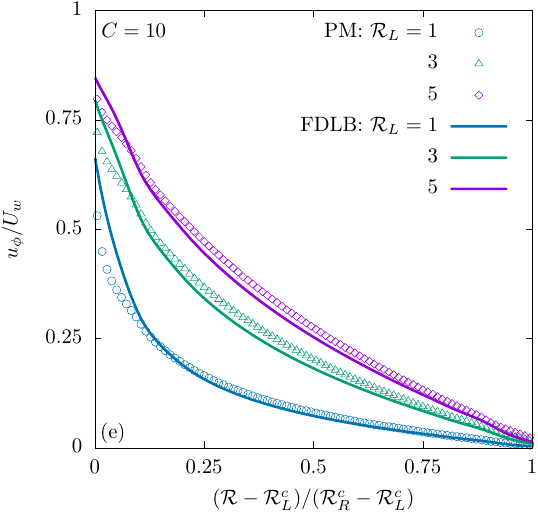}&
\includegraphics[width=0.325\linewidth]{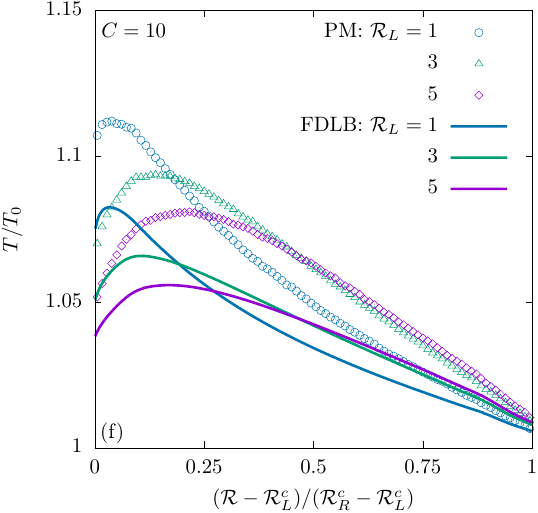}
\end{tabular}
\caption{Cylindrical Couette flow: Normalized (a,d) reduced density $\eta/\eta_0$, (b,e) azimuthal velocity $u_\phi/U_w$ and (c,f) temperature $T/T_0$ for a mean reduced density of $\eta_0=0.2$, two values of the confinement ratio and three values of the inner cylinder radius $\mathcal{R}_L\in\{1,3,5\}$. \label{fig:couettecyl_varRl_eta02}}
\end{figure*}

\begin{figure*}
\begin{tabular}{ccc}
\includegraphics[width=0.325\linewidth]{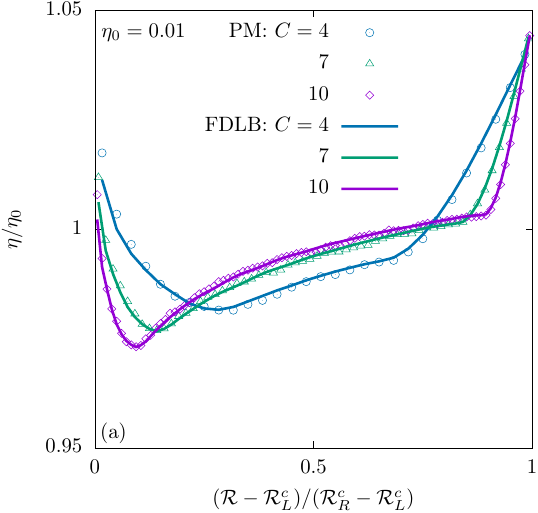}&
\includegraphics[width=0.325\linewidth]{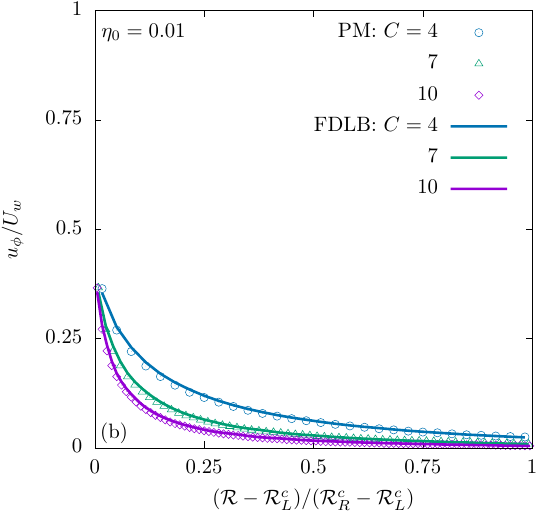}&
\includegraphics[width=0.325\linewidth]{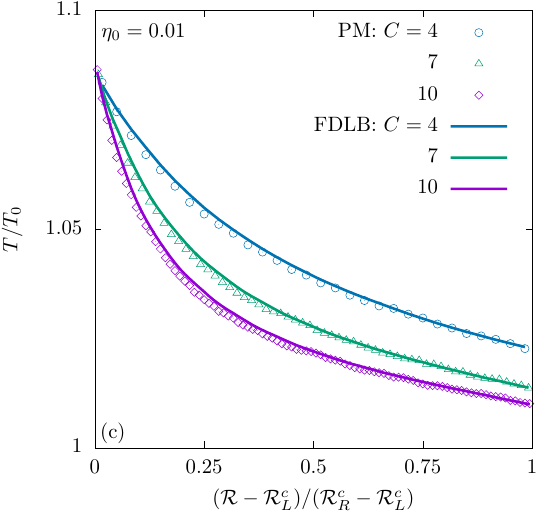}\\
\includegraphics[width=0.325\linewidth]{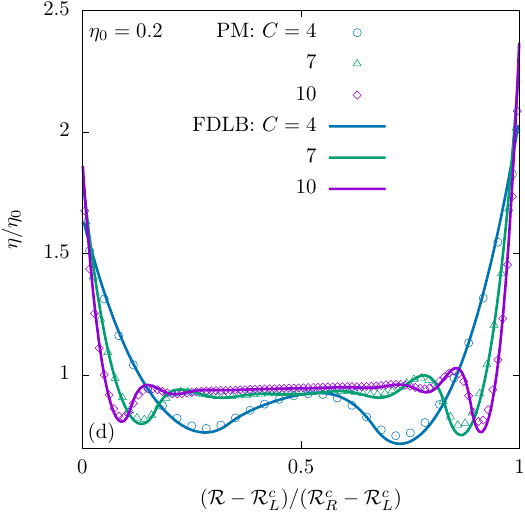}&
\includegraphics[width=0.325\linewidth]{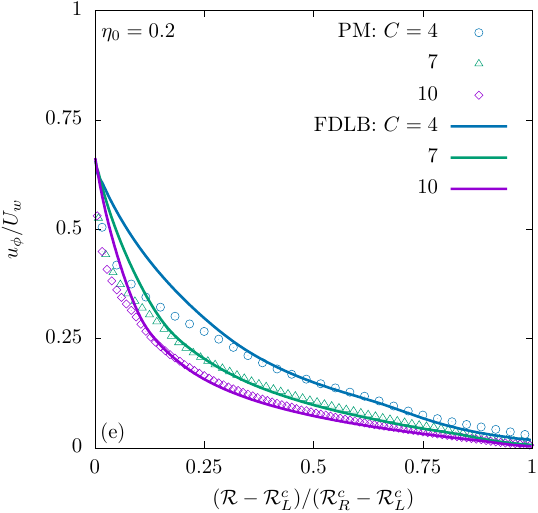}&
\includegraphics[width=0.325\linewidth]{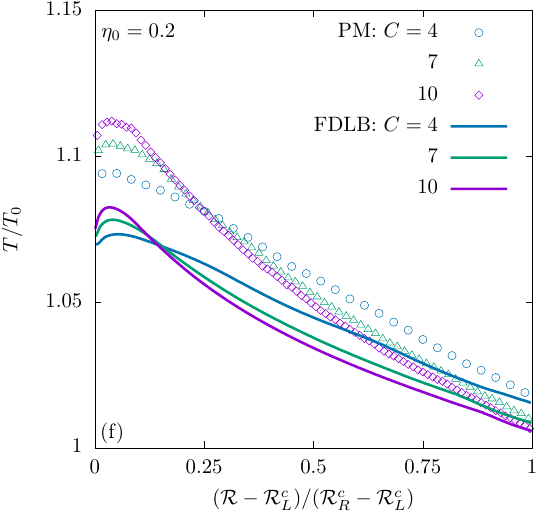}
\end{tabular}
\caption{Cylindrical Couette flow: Normalized (a,c) reduced density $\eta/\eta_0$, (b,e) azimuthal velocity $u_\phi/U_w$ and (c,f) temperature $T/T_0$ for the inner cylinder radius $\mathcal{R}_L=1$, two values of the mean reduced density $\eta_0\in\{0.01,0.2\}$ and three values of the confinement ratio $C\in\{4,7,10\}$. \label{fig:couettecyl_varR_rl1}}
\end{figure*}

In this subsection, we analyze the cylindrical Couette flow of a gas confined between two infinite coaxial cylinders. The inner cylinder rotates along the $\varphi$ axis with fixed velocity $U_w=\Omega_w R_L=\sqrt{k_B T_w/m}$, while the outer cylinder is at rest. The wall temperature of both cylinders is equal and constant ($T_w=T_0=1$). The simulations were conducted for three values of the mean reduced density $\eta_0\in\{0.01,0.1,0.2\}$, three values of the confinement ratio $C\in\{4,7,10\}$, corresponding to $\mathcal{L}_c\in\{3,6,9\}$ and three values of the inner cylinder radius $\mathcal{R}_L=\{1,3,5\}$ (with the outer cylinder radius $\mathcal{R}_R=\mathcal{R}_L+C$). The corresponding values of the Knudsen number associated with each case are summarized in Table~\ref{tab:quad}. The quadratures used in these simulations were HHLB$(7,100)\times$HLB$(7,8)$ for $\eta_0=0.01$ and HHLB$(7,20)\times$HLB$(7,8)$ for $\eta_0\in\{0.1,0.2\}$. Larger values of the quadrature orders do not bring any significant numerical changes to the macroscopic quantities of interest.
The expansion order $N = 7$ of the equilibrium distribution function is sufficient to recover the necessary moments for the implementation of the Shakhov relaxation time approximation.

In Fig.~\ref{fig:couettecyl_rl1_vareta}(a-c) we plot the normalized density $\eta/\eta_0$, azimuthal velocity $u_\phi/U_w$ and temperature $T/T_0$ for three values of the mean reduced density $\eta_0\in\{0.01,0.1,0.2\}$. The radius of the inner cylinder is $\mathcal{R}_L=1$ and the confinement ratio was $C=4$. Firstly we observed the influence of the curved boundary on the layering effect near the walls, i.e. the lower value of the reduced fluid density next to the inner cylindrical wall than next to the outer cylinder wall. As explained in  Appendix \ref{app:stationary}, when a fluid particle is located at a distance less than $\sigma$ from the boundary, a portion of its surface remains protected from collisions since there is not enough space available for a second particle to occupy that region. On the other hand, when curved boundaries are considered, this surface shrinks and the space available at a distance $\sigma$ from the center of the particle increases. It is worth remarking in Fig.~\ref{fig:couettecyl_rl1_vareta}(a) the nice match between the particle method PM results and the FDLB results for the reduced density profile. Both the azimuthal velocity and the temperature profiles obtained using FDLB are in good agreement with the PM results for small values of $\eta_0$.  When $\eta_0$ is increased the FDLB results start to depart from the PM results near the inner cylinder. Moving to a larger confinement ratio $C=10$, one can observe in Fig.~\ref{fig:couettecyl_rl1_vareta}(d-f) that the layering is more pronounced and the left and right layers are separated (do not interfere). The results show good agreement between FDLB and PM results. The discrepancies in azimuthal velocity are located near the inner cylinder and cover a region of size $\sigma$ ($1/9\mathcal{L}_c$ in this case), while the FDLB and the PM results temperature profiles are in qualitative agreement and differ only with a relative error of less than $5\%$.

Next, we focus on varying the inner cylinder radius $\mathcal{R}_L$. In Fig. \ref{fig:couettecyl_varRl_eta02} we plot the numerical results for the highest mean reduced density $\eta_0=0.2$, for a confinement ratio of $C=4$ in the first row and $C=10$ in the second row. The columns are dedicated to (from the left) the normalized reduced density $\eta/\eta_0$, the normalized azimuthal velocity $u_\phi/U_w$, and the normalized temperature $T/T_0$, respectively. One can immediately observe the change in the layering next to the inner cylinder as $R_L$ increases and the curvature diminishes. The FDLB results match very well the PM results even for this high mean reduced density. The azimuthal velocity preserves a discrepancy next to the inner cylinder which is also present in the planar wall case to which these results tend as $R_L$ increases. The temperature has the same qualitative behavior in the FDLB results as the PM results but underestimates them by less than $5\%$.

Lastly, one can fix the inner cylinder radius at the highest curvature, i.e. $R_L=1$, and vary the mean reduced density $\eta_0$ and the confinement ratio $C$. In Fig.~\ref{fig:couettecyl_varR_rl1} we present the comparison of the results obtained using the FDLB and PM methods for $\eta_0=0.01$ and $\eta_0=0.2$ while varying the confinement ratio $C\in\{4,7,10\}$. The first column plots the normalized density and since we use a reduced coordinate $(\mathcal{R}-\mathcal{R}_L^c)/(\mathcal{R}_R^c-\mathcal{R}_L^c)$ the relative size of the particles diminishes and the layering is more pronounced. Excellent agreement is obtained for all values tested. In the second and third columns we represent the normalized azimuthal velocity and temperature and we obtained excellent agreement for a reduced density of $\eta_0=0.01$ as in this case the dense gas effect is small and we recover results close to the Boltzmann limit, as in Ref. \cite{BA19}. For $\eta_0=0.2$ we still have a nice agreement overall.

\begin{figure*}
\begin{tabular}{ccc}
\includegraphics[width=0.325\linewidth]{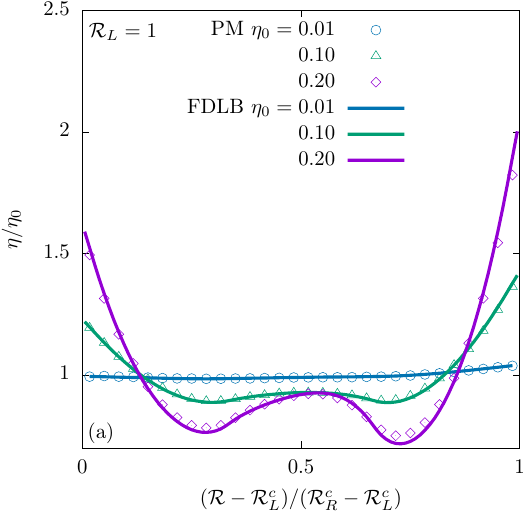}&
\includegraphics[width=0.325\linewidth]{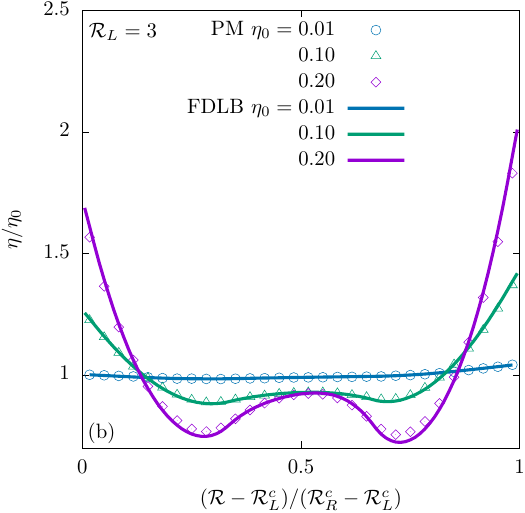}&
\includegraphics[width=0.325\linewidth]{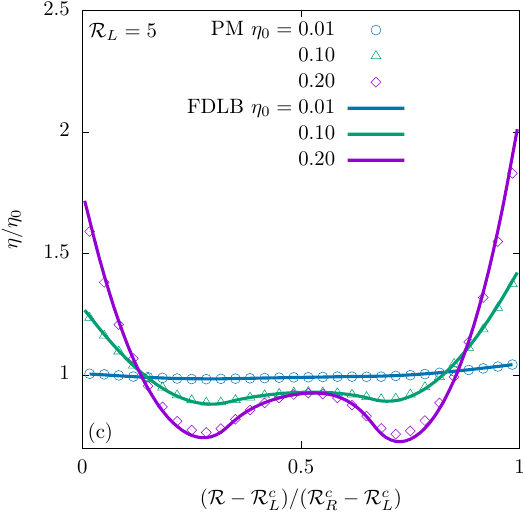}\\
\includegraphics[width=0.325\linewidth]{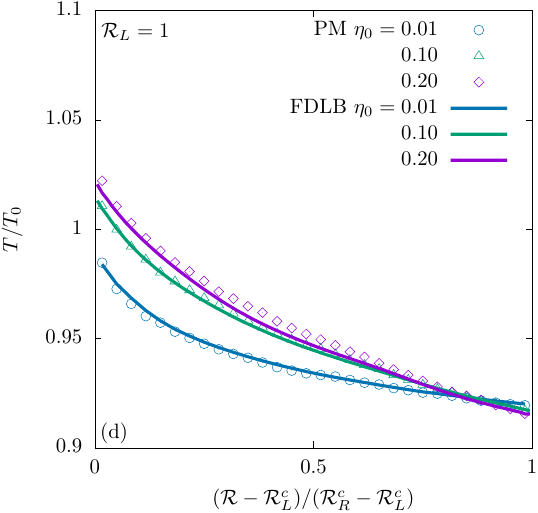}&
\includegraphics[width=0.325\linewidth]{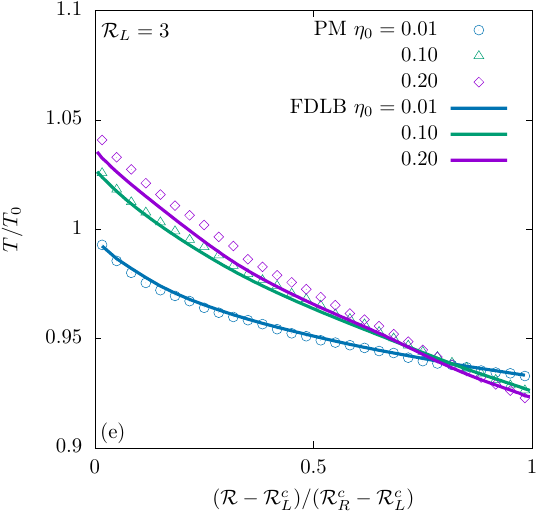}&
\includegraphics[width=0.325\linewidth]{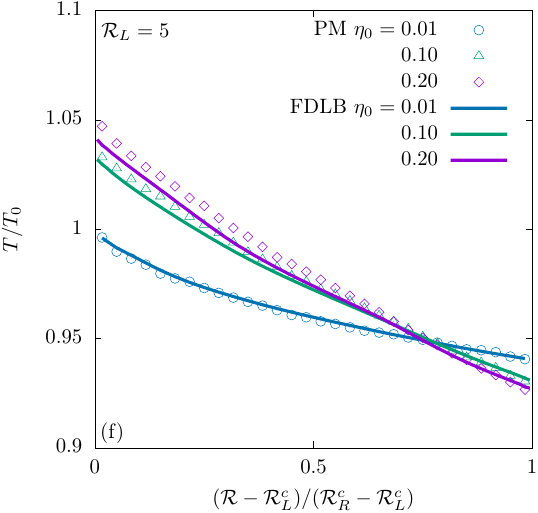}
\end{tabular}
\caption{Cylindrical Fourier flow: Normalized reduced density $\eta/\eta_0$ (a-c) and normalized temperature $T/T_0$(d-f), for $\mathcal{R}_L\in\{1,3,5\}$ and the confinement ratio $C=4$, while lines (FDLB) and points (PM) correspond to varying values of the mean reduced density $\eta_0\in\{0.01.0.1,0.2\}$.  \label{fig:termtrans_vareta}}
\end{figure*}

\begin{figure*}
\begin{tabular}{ccc}
\includegraphics[width=0.325\linewidth]{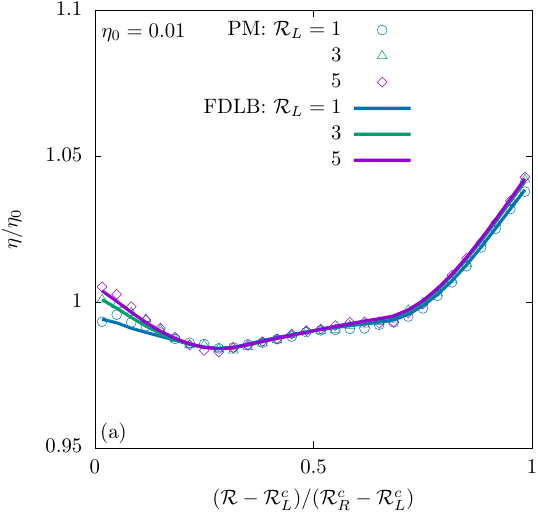}&
\includegraphics[width=0.325\linewidth]{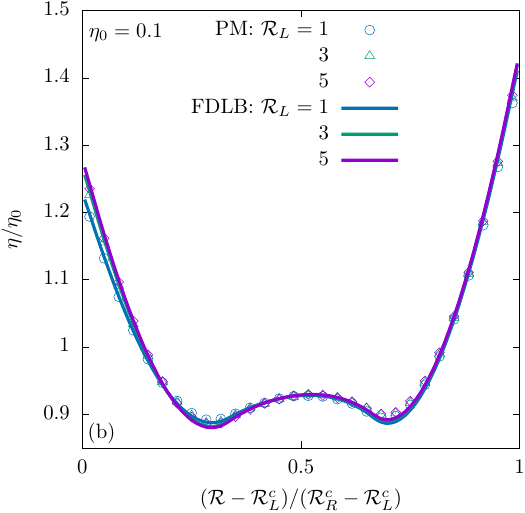}&
\includegraphics[width=0.325\linewidth]{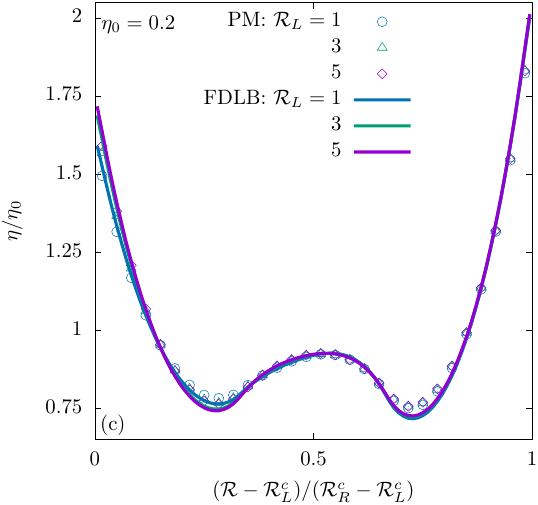}\\
\includegraphics[width=0.325\linewidth]{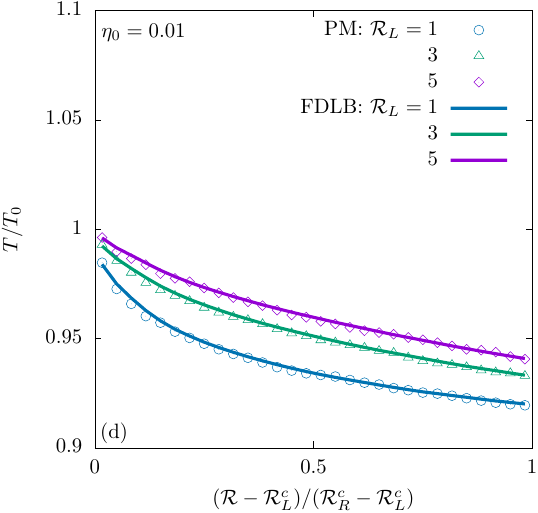}&
\includegraphics[width=0.325\linewidth]{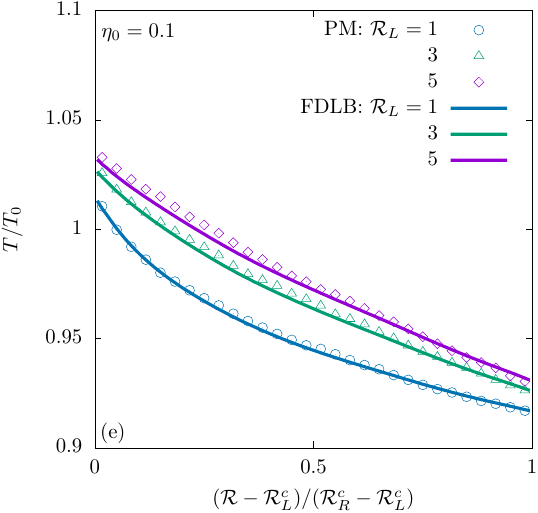}&
\includegraphics[width=0.325\linewidth]{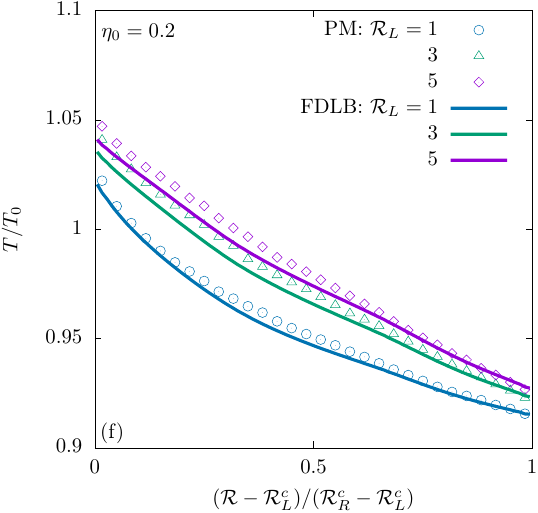}

\end{tabular}
\caption{Cylindrical Fourier flow: Normalized reduced density $\eta/\eta_0$ (a-c) and normalized temperature $T/T_0$(d-f), for $C=4$ and $\eta_0\in\{0.01.0.1,0.2\}$, while the lines (FDLB) and points (PM) correspond to varying values of the inner cylinder radius $\mathcal{R}_L\in\{1,3,5\}$.  \label{fig:termtrans_varRl} }
\end{figure*}

\begin{figure*}
\begin{tabular}{ccc}
\includegraphics[width=0.325\linewidth]{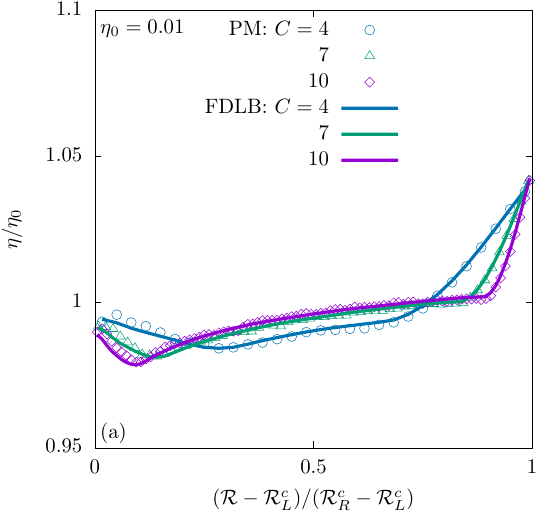}&
\includegraphics[width=0.325\linewidth]{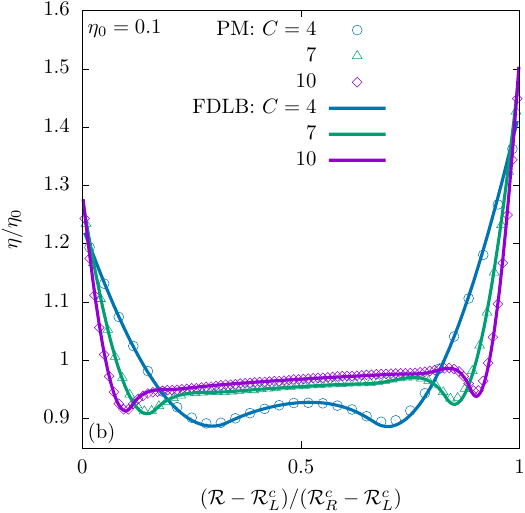}&
\includegraphics[width=0.325\linewidth]{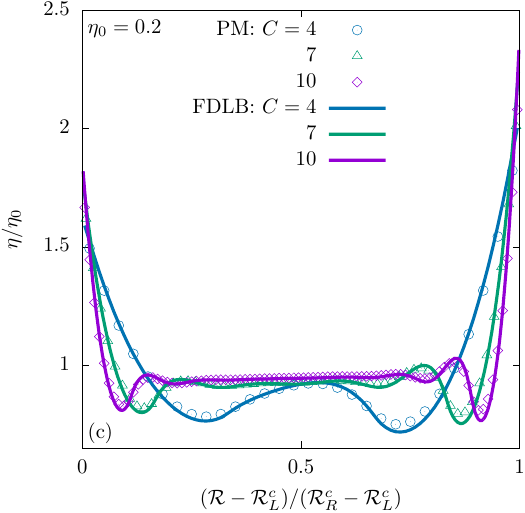}\\
\includegraphics[width=0.325\linewidth]{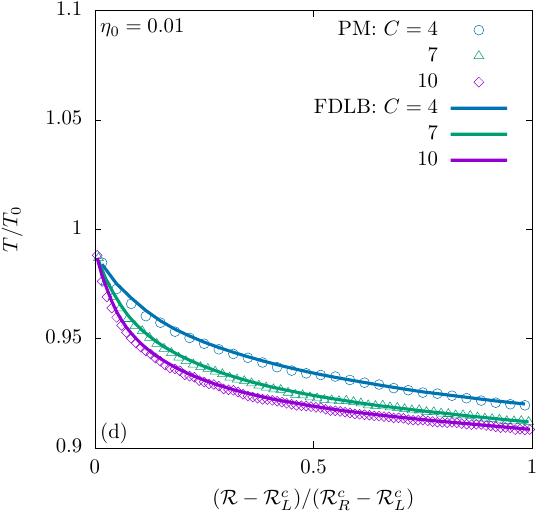}&
\includegraphics[width=0.325\linewidth]{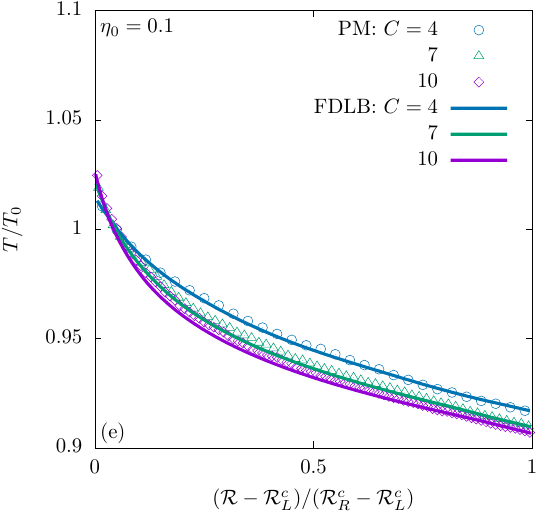}&
\includegraphics[width=0.325\linewidth]{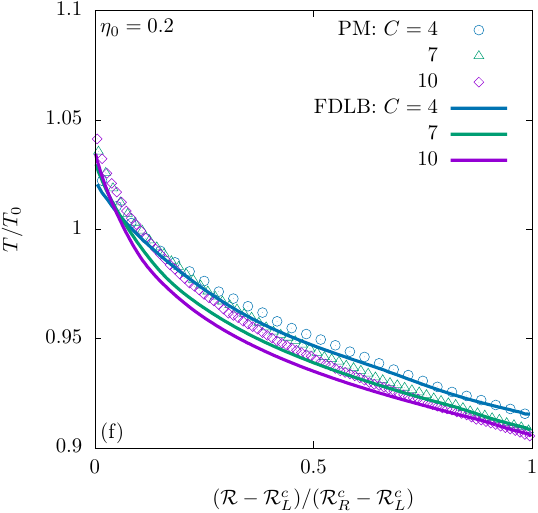}
\end{tabular}
\caption{Cylindrical Fourier flow: Normalized reduced density $\eta/\eta_0$ (a-c) and normalized temperature $T/T_0$(d-f), for $\mathcal{R}_L=1$ and $\eta_0\in\{0.01.0.1,0.2\}$, while the lines (FDLB) and points (PM) correspond to varying values of the confinement ratio $C\in\{4,7,10\}$.  \label{fig:termtrans_varR} }
\end{figure*}

\subsection{Cylindrical and spherical Fourier flow}

In this section, we study the Fourier flow between coaxial cylinders and concentric spheres. The simulations were conducted for three values of the mean reduced density $\eta_0\in\{0.01,0.1,0.2\}$, three values of the confinement ratio $C\in\{4,7,10\}$, corresponding to $\mathcal{L}_c\in\{3,6,9\}$ and three values of the inner cylinder/sphere radius $\mathcal{R}_L=\{1,3,5\}$ (with the outer cylinder/sphere radius of $\mathcal{R}_R=\mathcal{R}_L+C$). The corresponding Knudsen numbers associated with these systems are summarized in Table~\ref{tab:quad}. The quadratures used in the cylindrical Fourier flow were HHLB$(7,60)\times$HLB$(7,8)$ for $\eta_0=0.01$ and HHLB$(7,20)\times$HLB$(7,8)$ for $\eta_0\in\{0.1,0.2\}$, while for the spherical case a mixed quadrature of HHLB$(7,60)\times$HLB$(7,8)\times$HLB$(7,8)$ for $\eta_0=0.01$ and HHLB$(7,20)\times$HLB$(7,8)\times$HLB$(7,8)$ for $\eta_0\in\{0.1,0.2\}$. Larger values of the quadrature orders do not bring any significant numerical changes to the macroscopic quantities of interest.

\subsubsection{Cylindrical Fourier flow}

Here we analyze the Fourier flow in a gas confined between two infinite coaxial cylinders. The inner cylinder temperature is fixed at $T_L=T_0+\Delta T$ and on the outer cylinder the temperature is $T_R=T_0-\Delta T$, with $T_0=1$, as presented in Fig.~\ref{fig:setups}(b).
The results are grouped by varying the mean reduced density $\eta_0$ in Fig.~\ref{fig:termtrans_vareta}, the inner cylinder radius $\mathcal{R}_L$ in Fig.~\ref{fig:termtrans_varRl} and the confinement ratio $C$ in Fig.~\ref{fig:termtrans_varR}.

As observed in Fig.~\ref{fig:termtrans_vareta}, as the mean reduced density increases the layering is more pronounced in the reduced density plots (top row), while the temperature profile has some discrepancies but is overall in good agreement. As the inner cylinder radius $\mathcal{R}_L$ is increased the reduced density on the inner cylinder increases as it tends towards the planar wall results \cite{BS24}, and the temperature tends towards a more linear profile as the curvature is diminished. This is more evident in Fig.~\ref{fig:termtrans_varRl}, where we vary the inner radius while keeping the mean reduced density and the confinement ratio fixed. For small $\eta_0$ we obtain excellent agreement between the FDLB and PM approaches for both density and temperature, while for the larger values of $\eta_0$ the differences in temperature profiles are negligible, at around $2-3\%$.

In Fig.~\ref{fig:termtrans_varR} we plot the results obtained by varying the confinement ratio $C$ while keeping fixed the mean reduced density $\eta_0$ and the inner cylinder radius $\mathcal{R}_L$. The layering in the density profile is changing due to the use of the reduced coordinate $(\mathcal{R}-\mathcal{R}_L^c)/(\mathcal{R}_R^c-\mathcal{R}_L^c)$, such that the relative size of the particle is smaller. Excellent agreement is observed for all values of the confinement ratio at $\eta_0=0.01$, while there are some discrepancies at larger values of the mean reduced density, albeit not significant.

\subsubsection{Spherical Fourier flow}

\begin{figure*}
\begin{tabular}{ccc}
\includegraphics[width=0.325\linewidth]{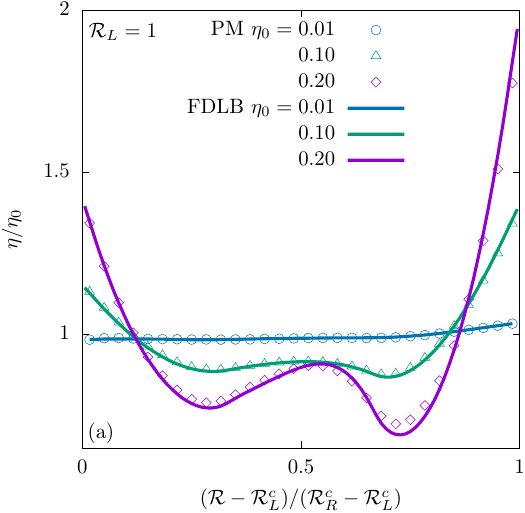}&
\includegraphics[width=0.325\linewidth]{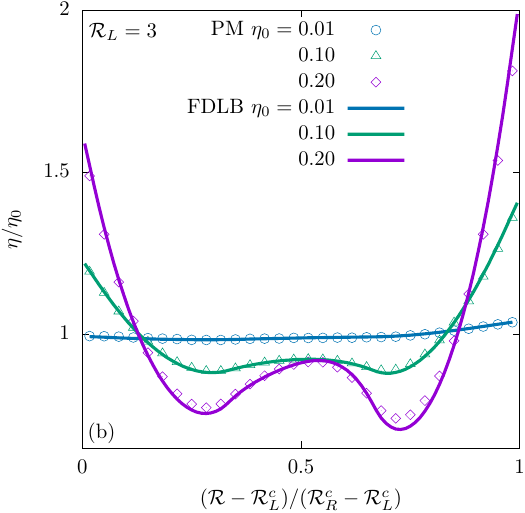}&
\includegraphics[width=0.325\linewidth]{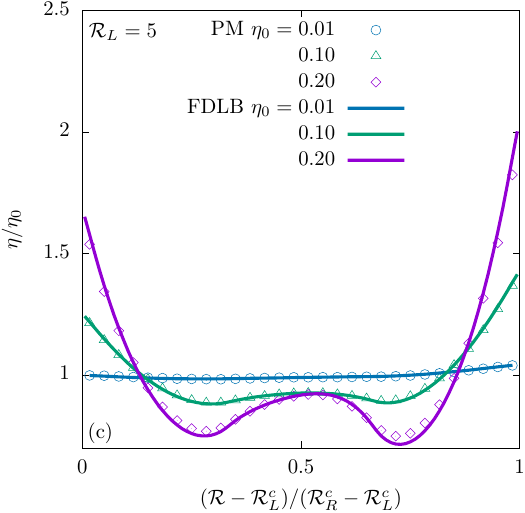}\\
\includegraphics[width=0.325\linewidth]{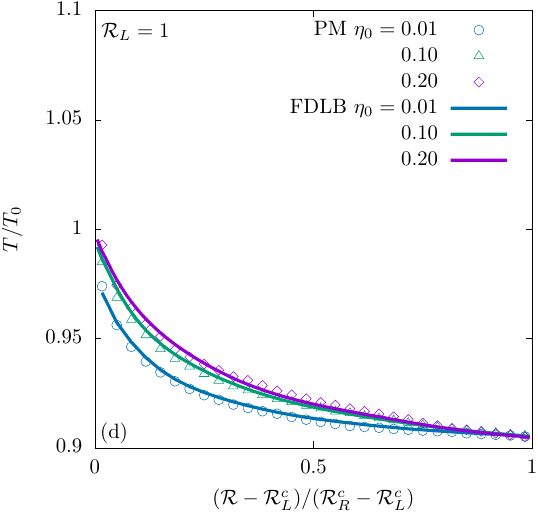}&
\includegraphics[width=0.325\linewidth]{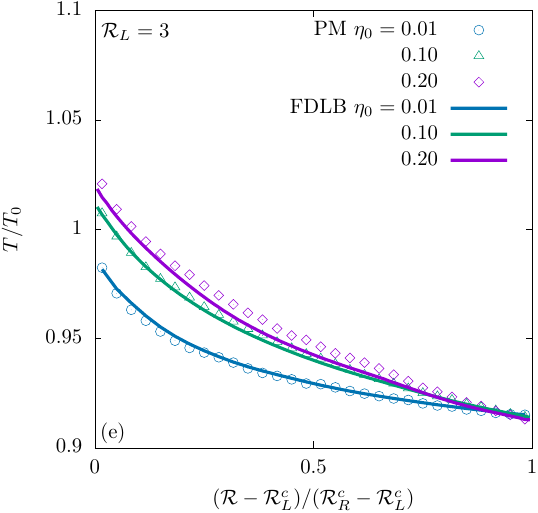}&
\includegraphics[width=0.325\linewidth]{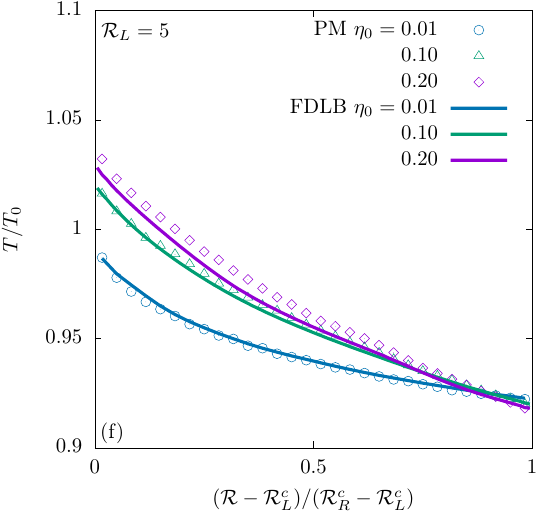}
\end{tabular}
\caption{Spherical Fourier flow: Normalized reduced density $\eta/\eta_0$ (a-c) and normalized temperature $T/T_0$(d-f) for $\mathcal{R}_L\in\{1,3,5\}$, and a confinement ratio of $C=4$, while lines (FDLB) and points (PM) correspond to varying values of the mean reduced density $\eta_0\in\{0.01.0.1,0.2\}$.  \label{fig:termtrans_vareta_sph}}
\end{figure*}

\begin{figure*}
\begin{tabular}{ccc}
\includegraphics[width=0.325\linewidth]{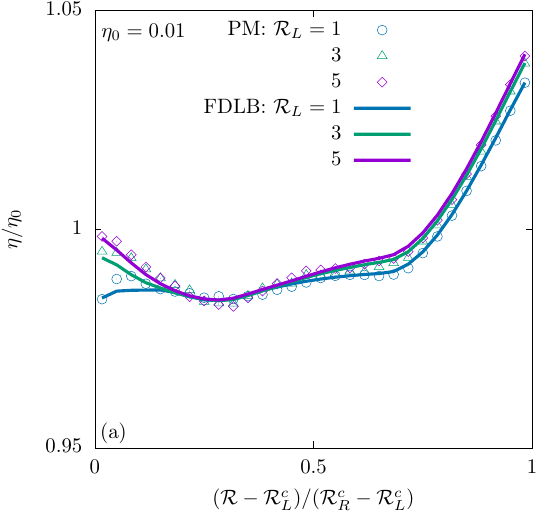}&
\includegraphics[width=0.325\linewidth]{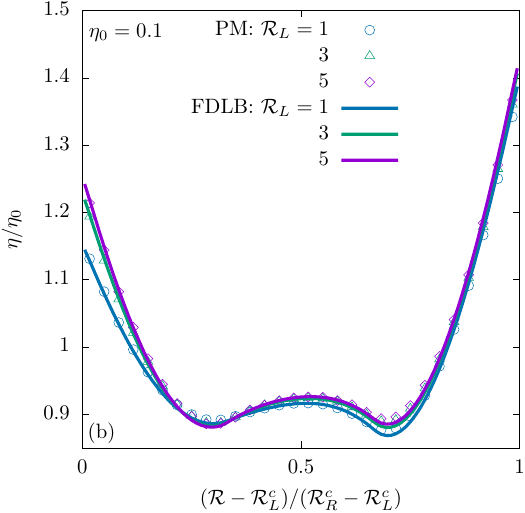}&
\includegraphics[width=0.325\linewidth]{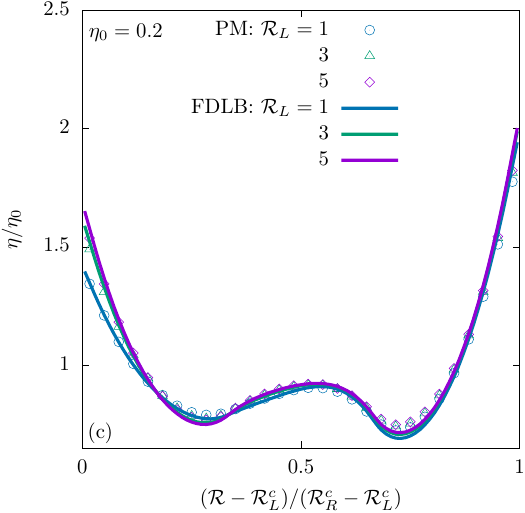}\\
\includegraphics[width=0.325\linewidth]{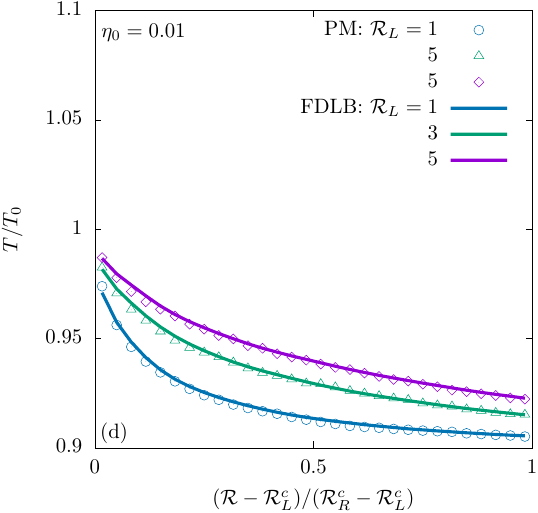}&
\includegraphics[width=0.325\linewidth]{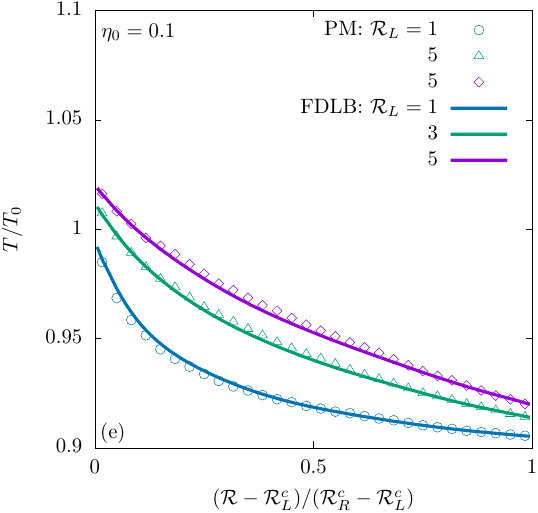}&
\includegraphics[width=0.325\linewidth]{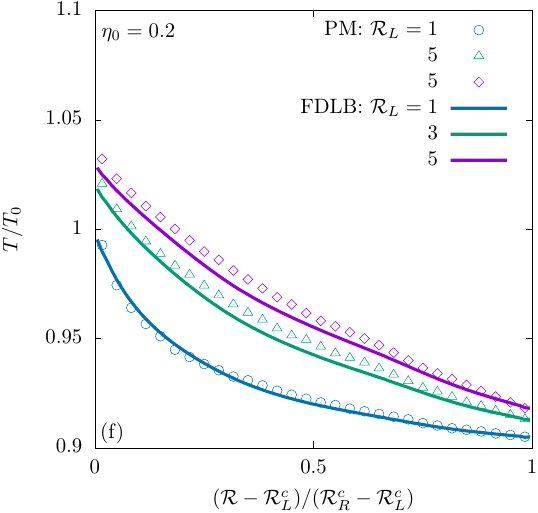}

\end{tabular}
\caption{Spherical Fourier flow: Normalized reduced density $\eta/\eta_0$ (a-c) and normalized temperature $T/T_0$(d-f) for $\eta_0\in\{0.01.0.1,0.2\}$, and a confinement ratio of $C=4$, while the lines (FDLB) and points (PM) correspond to varying values of the inner sphere radius $\mathcal{R}_L\in\{1,3,5\}$.  \label{fig:termtrans_varRl_sph}  }
\end{figure*}

\begin{figure*}
\begin{tabular}{ccc}
\includegraphics[width=0.325\linewidth]{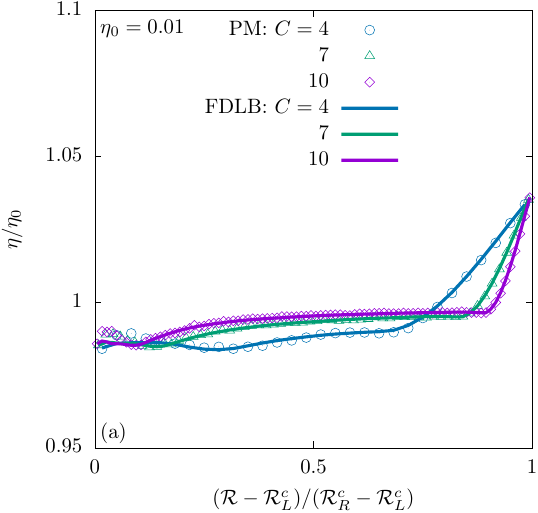}&
\includegraphics[width=0.325\linewidth]{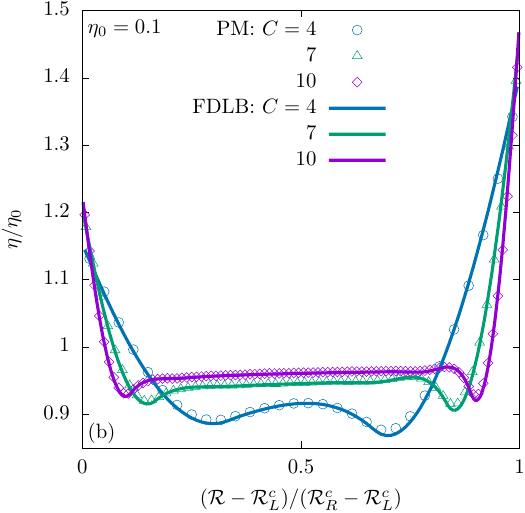}&
\includegraphics[width=0.325\linewidth]{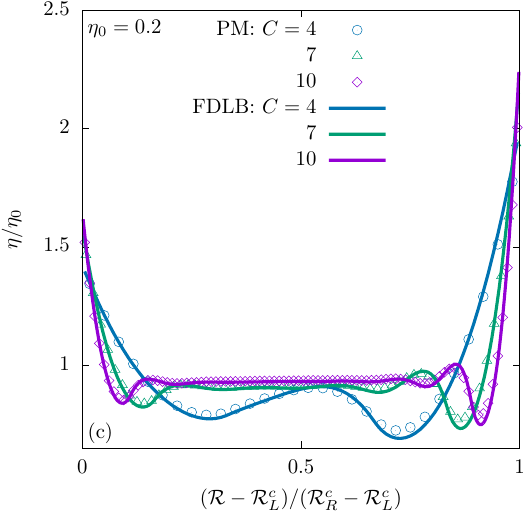}\\
\includegraphics[width=0.325\linewidth]{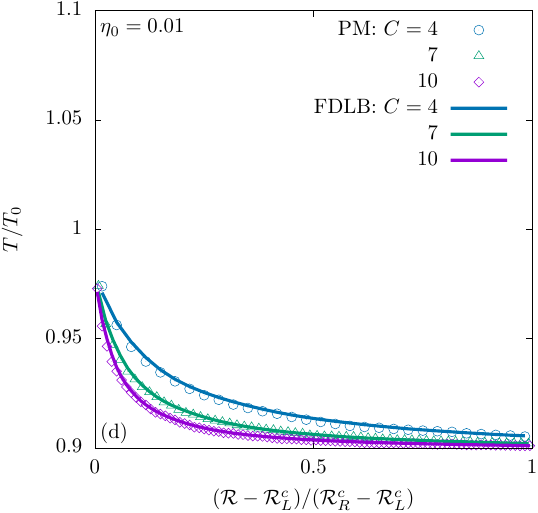}&
\includegraphics[width=0.325\linewidth]{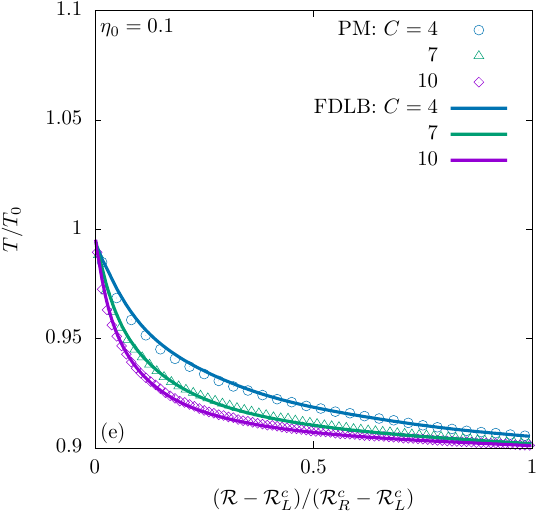}&
\includegraphics[width=0.325\linewidth]{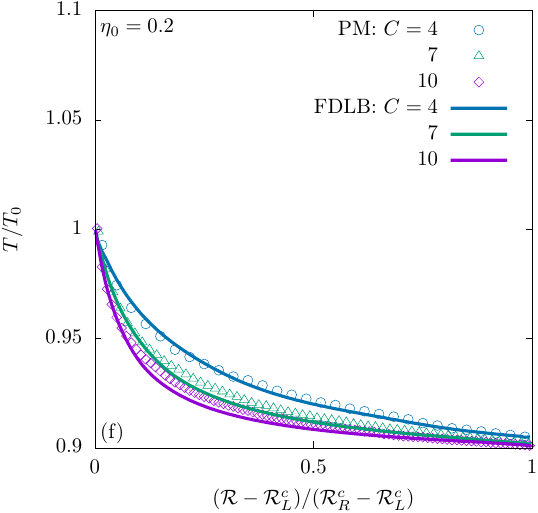}
\end{tabular}
\caption{Spherical Fourier flow: Normalized reduced density $\eta/\eta_0$ (top row) and normalized temperature $T/T_0$ for $\mathcal{R}_L=1$ and $\eta_0\in\{0.01.0.1,0.2\}$. The lines (FDLB) and points (PM) correspond to varying values of the confinement ratio $C\in\{4,7,10\}$.  \label{fig:termtrans_varR_sph} }
\end{figure*}

In this section, we analyze the Fourier flow in a gas confined between two concentric spheres. The inner sphere temperature is fixed at $T_L=T_0+\Delta T$ and on the outer sphere the temperature is $T_R=T_0-\Delta T$, with $T_0=1$, as presented in Fig.~\ref{fig:setups}(c). As in the case of the cylindrical Fourier flow, the results are grouped by varying the mean reduced density $\eta_0$ in Fig.~\ref{fig:termtrans_vareta_sph}, the inner sphere radius $\mathcal{R}_L$ in Fig.~\ref{fig:termtrans_varRl_sph} and the confinement ratio $C$ in Fig.~\ref{fig:termtrans_varR_sph}.

The results obtained by varying the mean reduced density $\eta_0$ and the inner sphere radius $\mathcal{R}_L$  are compiled in Fig.~\ref{fig:termtrans_vareta_sph}. The increase in the mean reduced density leads to a more pronounced layering in the reduced density profiles (top row), while the temperature profile has some discrepancies but is overall in good agreement. As the inner sphere radius $\mathcal{R}_L$ is increased the reduced density on the inner sphere increases as it tends towards the planar wall results \cite{BS24}, and the temperature tends towards a more linear profile as the curvature is diminished. This is more evident in Fig.~\ref{fig:termtrans_varRl_sph}, where we vary the inner radius while keeping the mean reduced density and the confinement ratio fixed. We obtain excellent agreement between the FDLB and PM approaches for both density and temperature for small mean reduced density $\eta_0$, while for the larger values of $\eta_0$ the differences in temperature are negligible, at around $2-3\%$.

The results obtained by varying the confinement ratio $C$ while keeping fixed the mean reduced density $\eta_0$ and the inner sphere radius $\mathcal{R}_L$ are plotted in Fig.~\ref{fig:termtrans_varR_sph}. As before, due to the use of the reduced coordinate $(\mathcal{R}-\mathcal{R}_L^c)/(\mathcal{R}_R^c-\mathcal{R}_L^c)$ the layering in the reduced density profile is modified since the relative size of the particle is smaller. Excellent agreement is observed for all values of the confinement ratio at $\eta_0=0.01$, while there are insignificant discrepancies at larger values of the mean reduced density.

\subsubsection{Geometry comparisons}

\begin{figure*}
\begin{tabular}{ccc}
\includegraphics[width=0.325\linewidth]{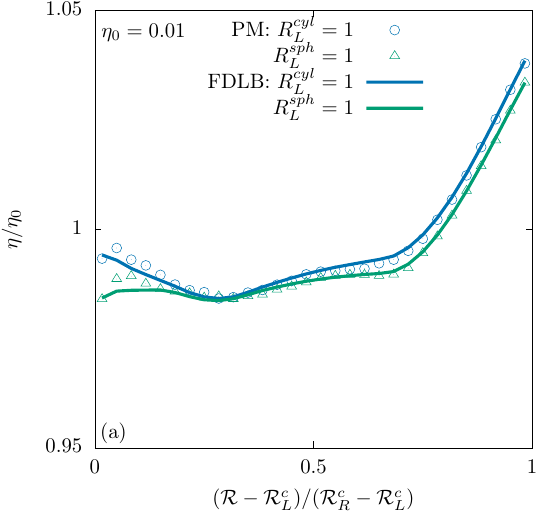}&
\includegraphics[width=0.325\linewidth]{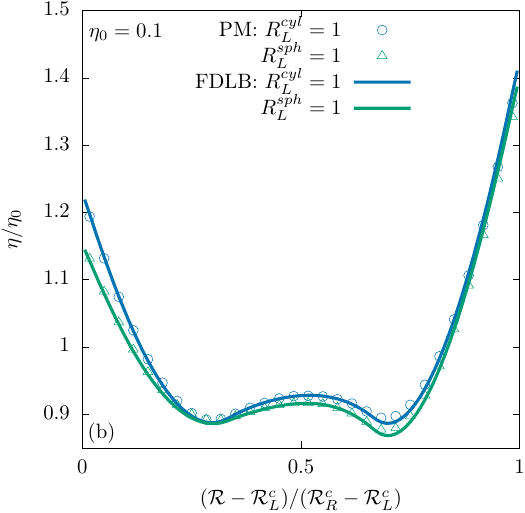}&
\includegraphics[width=0.325\linewidth]{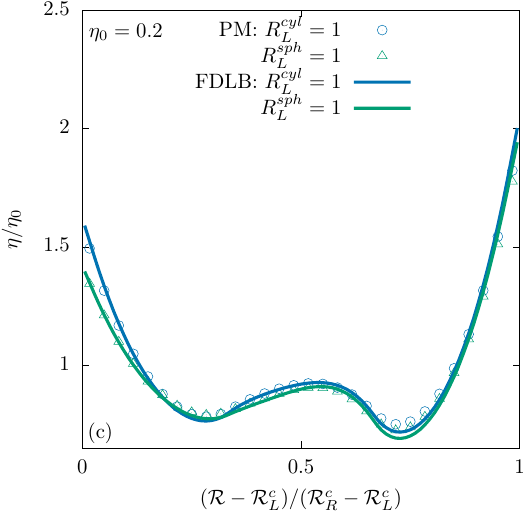}\\
\includegraphics[width=0.325\linewidth]{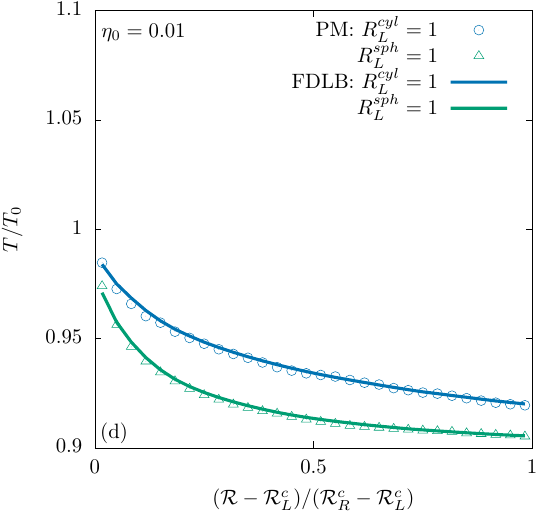}&
\includegraphics[width=0.325\linewidth]{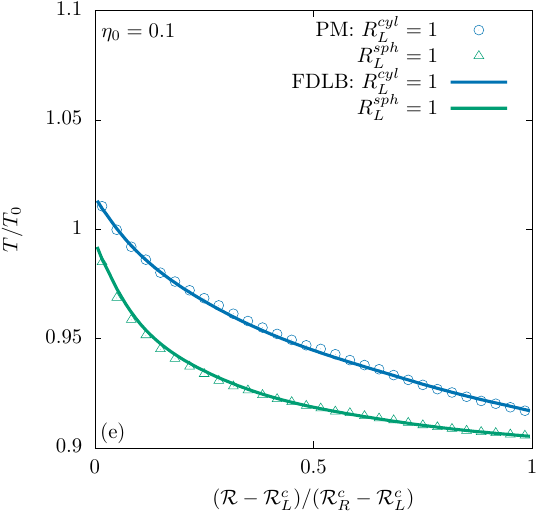}&
\includegraphics[width=0.325\linewidth]{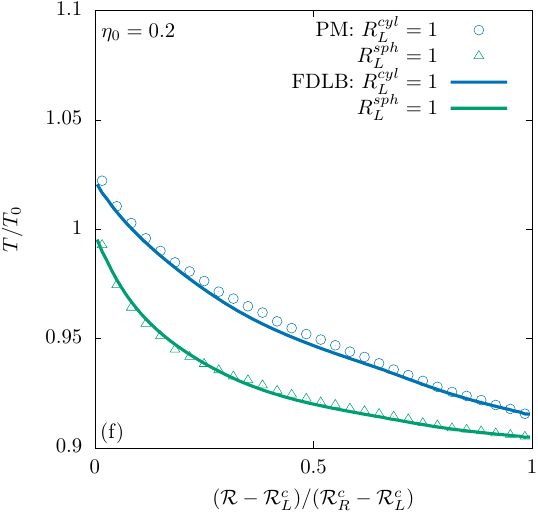}
\end{tabular}
\caption{Fourier flow: Normalized reduced density $\eta/\eta_0$(a-c) and temperature $T/T_0$(d-f) for the confinement ratio $C=4$, the inner cylinder/sphere radius $\mathcal{R}_L=1$, three values of the mean reduced density $\eta_0\in\{0.01.0.1,0.2\}$ and the two geomtries, cylindrical and spherical. The variation in the surface of the particle protected from collisions is clearly observed in both reduced density and temperature, with the temperature profile in spherical geometry being always below the cylindrical one. \label{fig:termtrans_varcoord_R4}}
\end{figure*}

\begin{figure*}
\begin{tabular}{ccc}
\includegraphics[width=0.325\linewidth]{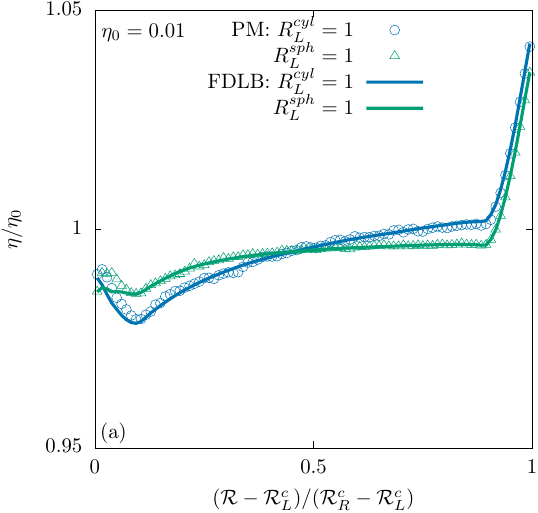}&
\includegraphics[width=0.325\linewidth]{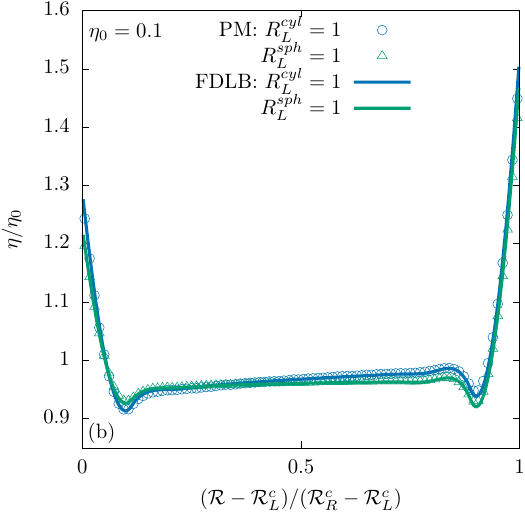}&
\includegraphics[width=0.325\linewidth]{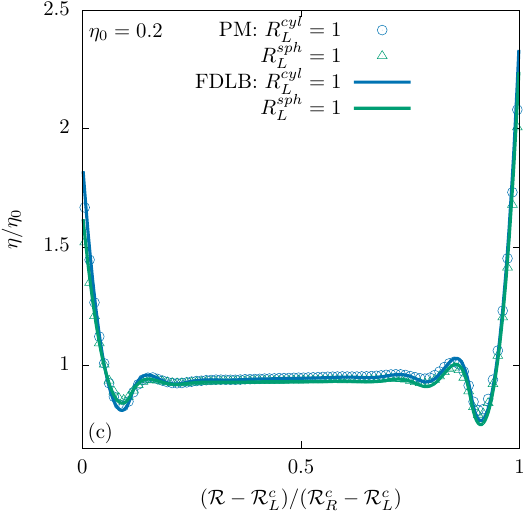}\\
\includegraphics[width=0.325\linewidth]{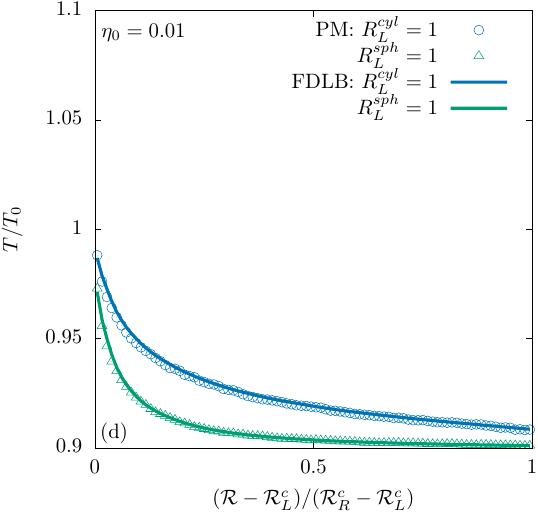}&
\includegraphics[width=0.325\linewidth]{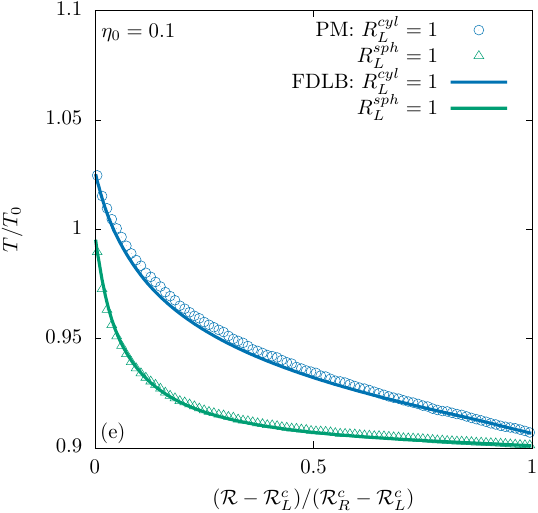}&
\includegraphics[width=0.325\linewidth]{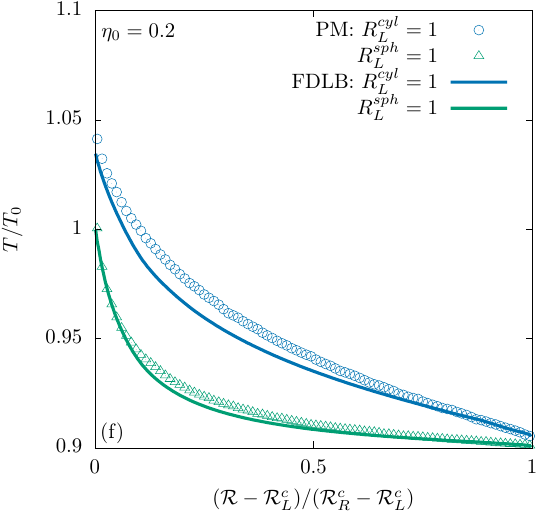}
\end{tabular}
\caption{Fourier flow: Normalized reduced density $\eta/\eta_0$(a-c) and temperature $T/T_0$(d-f) for the confinement ratio $C=10$. The rest of parameters are the same as Fig.\ref{fig:termtrans_varcoord_R4}. \label{fig:termtrans_varcoord_R10}}
\end{figure*}

In this subsection, we directly compare the results of the two geometries used in the case of the Fourier flow. In Figs.~\ref{fig:termtrans_varcoord_R4} and \ref{fig:termtrans_varcoord_R10} we plot the normalized reduced density $\eta/\eta_0$(a-c) and temperature $T/T_0$ (d-f), when the inner cylinder/sphere radius is $\mathcal{R}_L=1$, for three values of the mean reduced density $\eta_0\in\{0.01.0.1,0.2\}$ and both geometries considered (cylindrical and spherical), for a confinement ratio of $C=4$ and $C=10$, respectively. The direct comparison shows the extent to which the variation in the volume excluded by the boundary affects the layering near the wall, especially near the inner cylindrical/spherical boundary, when looking at the reduced density profile. When comparing the temperature profiles we can observe that the spherical profile is always below the cylindrical one. As expected the results obtained with the FDLB overlap very well with the PM results for low mean reduced density $\eta_0$ , and have a reasonable accuracy in the temperature profile.

Furthermore, one can compare the radial heat fluxes obtained in the two geometries. According to the conservation laws, the following quantities are constant through the channel in the cylindrically and spherically symmetric Fourier flows:
\begin{eqnarray}
\text{Cylindrical: }& q^R R=const.\\
\text{Spherical: }& q^r r^2=const.
\end{eqnarray}
In order to compare these values to the planar case we plot the following quantities with respect to the inner cylinder/sphere radius: $q^RR/\mathcal{R}_L,$ and $~q^r r^2/\mathcal{R}_L^2$. The results are summarized in Fig.~\ref{fig:termtrans_heatflux} for (a) $\eta_0=0.01$, (b) $\eta_0=0.1$ and (c) $\eta_0=0.2$, for both the FLDB and the PM results. The confinement ratio is fixed at $C=4$ and the inner cylinder/sphere radius is increased from $\mathcal{R}_L=1$ to $\mathcal{R}_L=50$. At low $\eta_0$ we have an excellent agreement throughout the range of inner cylinder/sphere radius, while as we increase the mean reduced density the value deviates as was the case for a planar wall, reported in Ref.~\cite{BS24}, which is represented in the figures as $\mathcal{R}_L\rightarrow\infty$. As observed in the planar case, the FDLB overestimates the PM results quite a lot, but this is expected for higher-order moments since the simplified Enskog collision operator approximation discards the higher-order contributions to the collisional momentum and energy transfer. The simulation results obtained using the PM contain the total heat flux, i.e. the sum of the kinetic and the potential contributions defined in Ref.~\cite{BS24}.

\begin{figure*}
\begin{tabular}{ccc}
\includegraphics[width=0.325\linewidth]{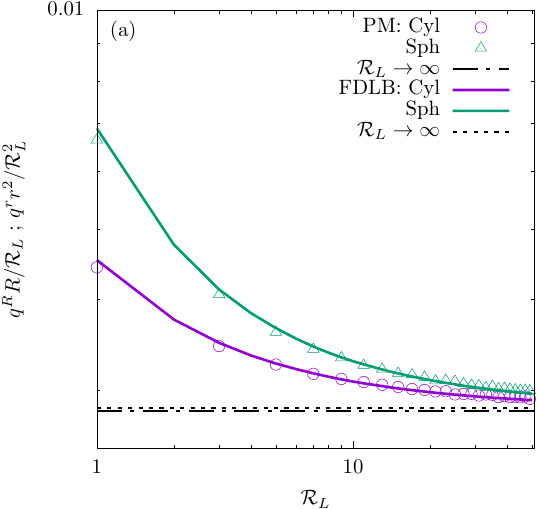}&
\includegraphics[width=0.325\linewidth]{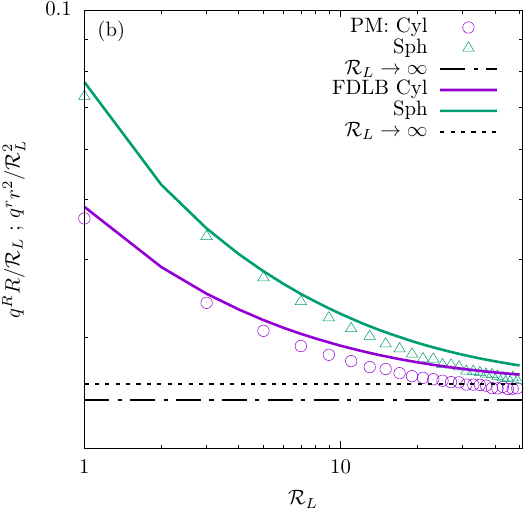}&
\includegraphics[width=0.325\linewidth]{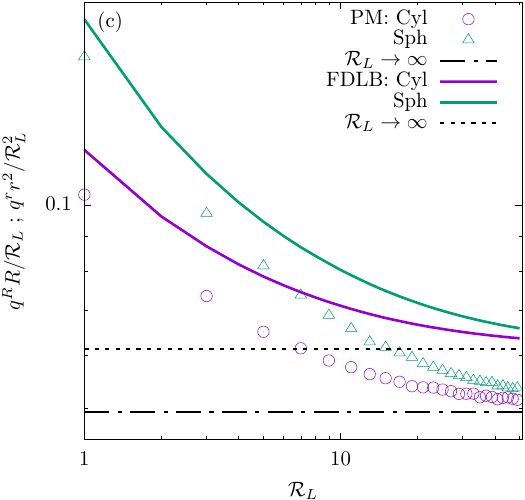}\\
(a)&(b)&(c)
\end{tabular}
\caption{Fourier flow: The contants $q^RR/\mathcal{R}_L,$ and $q^r r^2/\mathcal{R}_L^2$ for a temperature difference of $\Delta T=0.1$, confinement ratio of $C=4$, mean reduced density of $\eta_0$ of (a) $0.01$, (b) 0.1 and (c) 0.2, and variable inner cylinder/sphere radius $\mathcal{R}_L$. As the inner $\mathcal{R}_L$ increases, the two geometries approach a planar wall ($\mathcal{R}_L\rightarrow\infty$) and the results tend to match the results obtained in Ref.~\cite{BS24}, which are represented as dashed lines. \label{fig:termtrans_heatflux}}
\end{figure*}

\section{Conclusions\label{sec:conclusions}}

In this work, a series of dense gas flows bounded by curvilinear walls were considered in order to validate the proposed finite-difference Lattice Boltzmann model employing the simplified Enskog collision integral. For this purpose, the Enskog equation in curvilinear coordinates was written with respect to orthonormal vielbein fields (triads in $3D$), by extending the formalism introduced in Ref.~\cite{BA19} for the Boltzmann equation. The vielbein can be used to align the momentum space along the coordinate directions, while also decoupling the dependence of $(\vp - m\vu)^2$ appearing in the Maxwell-Boltzmann equilibrium distribution on the induced metric tensor. In the simplified Enskog collision model, the Enskog collision integral is approximated using a Taylor expansion and retaining the first-order gradients. Following Ref. \cite{BA19}, a half-range Gauss-Hermite quadrature was used on the axis normal to the curved boundaries studied. The model was benchmarked in three setups: cylindrical Couette and Fourier flows between coaxial cylinders, as well as the Fourier flow between concentric spheres.
The simulation parameters range from a low reduced density value ($\eta_0=0.01$) to a relatively high value ($\eta_0=0.2$), three values of the inner cylinder radius $\mathcal{R}_L\in\{1,3,5\}$ and three values of the confinement ratio $C\in\{4,7,10\}$.

The FDLB results obtained for the cylindrical Couette flow and the cylindrical and spherical Fourier flow were validated against the corresponding PM results. Reasonable agreement was observed throughout the parameter range. More specifically, our kinetic model adequately captures the effects of denseness, density inhomogeneity, as well as nonequilibrium phenomena within the range of flow parameters investigated. It is important to consider that when a fluid molecule is located at a distance less than a molecular diameter $\sigma$ from the wall, a portion of its surface remains protected from collisions since there is not sufficient space available for a second molecule to occupy that part of the spatial domain. As a result, the particle is pushed toward the wall. When dealing with a curved boundary this effect is either diminished or enhanced if a concave or convex boundary is involved. In our case, the more pronounced effect is on the inner cylinder/sphere where the available space is increased due to the shape of the boundary. As such the layering effect is inversely proportional to the curvature. This variation in the layering effect is well captured by the FDLB model proposed.

For the cylindrical Couette flow, we recover with good accuracy the density profile in the channel, while a reasonable accuracy is observed for the azimuthal velocity and temperature profile, with discrepancies between the FDLB and PM results being lower than $5\%$.

For the Fourier flow, we present the density and temperature distribution in the channel for all combinations of the input parameters. Good agreement is observed throughout the entire parameter range. We also include a comparison of the two geometries, pointing to their differences and the results which are in accordance with expectations based on the excluded volume for collision due to the variable curvature/geometry. Additionally, a discussion about the heat flux between concentric cylinders and concentric spheres is presented. Channel constants derived from heat flux show larger values for high curvature and reach asymptotically the planar case results as the radius of the cylinders/spheres is increased.

In conclusion, the presented model demonstrates its capability to handle curved geometries for moderately dense gases.
Moving forward, our goal is to implement the second-order terms in the Taylor expansion of the Enskog collision integral as well as attractive forces between molecules to address multiphase flows.

\appendix
\section{Volume intersections \label{appendix:volintersect}}

In order to evaluate the smoothed density $\bar{n}$ in Eq. \eqref{eq:fm}, in cylindrical coordinates we employ the procedure presented in Ref.\cite{LL90}. The intersection of the cylinder of radius $R$ and the sphere of radius $r$ is given by:
\begin{multline}
 V(r,R,b)=\frac{4\pi}{3}r^3\Theta(R-b)+\frac{4}{3\sqrt{A-C}}\left\{ \Pi(k,-\alpha^2)\frac{B^2s}{C}\right.\\ \left.
 +K(k)\left[s(A-2B)+(A-B)\frac{3B-C-2A}{3} \right]\right.\\\left.
 +E(k)(A-C)\left( -s+\frac{2A+2C-4B}{3} \right)
 \right\}, r<b+R
\end{multline}
where $b$ denotes the smallest distance of the axis of the cylinder to the center of the sphere and $\Theta$ is the Heaviside step function. We have limited to the description of the intersection volume to the case used in the present study when $r<R$.
In the above, we have introduced the elliptic integrals of the first, second, and third kind:
\begin{equation}
 K(k)\equiv\int_0^1\frac{dz}{\sqrt{1-z^2}\sqrt{1-k^2z^2}},\quad E(k)\equiv\int_0^1\frac{dz\sqrt{1-z^2}}{\sqrt{1-k^2z^2}}\nonumber
\end{equation}
\begin{equation}
 \Pi(k,-\alpha^2)\equiv\int_0^1\frac{dz}{(1-\alpha^2z^2)\sqrt{1-z^2}\sqrt{1-k^2z^2}}
\end{equation}
and the following quantities have been defined:
\begin{equation}
 A=\max\left(r^2,(b+R)^2 \right),\quad B=\min\left( r^2,(b+R)^2 \right),\quad C=(b-R)^2
\end{equation}
\begin{equation}
 k^2=\frac{B-C}{A-C},\quad -\alpha^2=\frac{B-C}{C},\quad s=(b+R)(b-R)
\end{equation}

These expressions were implemented through the use of the GNU Scientific Library (GSL)\cite{GNU}, a numerical library for C and C++ programmers.

In the spherical coordinates, one must evaluate the intersection of two spheres, with radii $r_1$ and $r_2$ and the distance between the centers of the spheres of $d$, for which we have an analytical expression:
\begin{equation}
 V(r_1,r_2,d)=\frac{\pi (r_1 + r_2 - d)^2 (d^2 + 2 d r_2 - 3 r_2^2 + 2 d r_1 + 6 r_1 r_2 - 3 r_1^2)}{12d}
\end{equation}

\section{Dense gas at rest between coaxial cylinders and concentric spheres\label{app:stationary}}

\begin{figure*}
\begin{tabular}{ccc}
\includegraphics[width=0.32\linewidth]{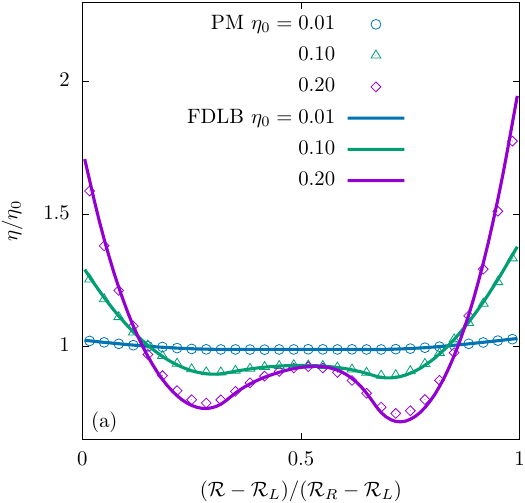}&
\includegraphics[width=0.32\linewidth]{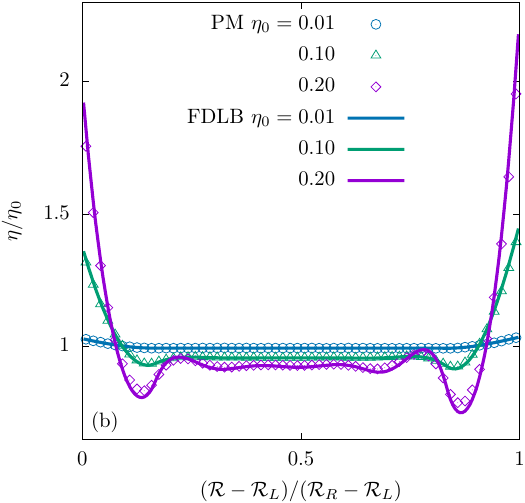}&
\includegraphics[width=0.32\linewidth]{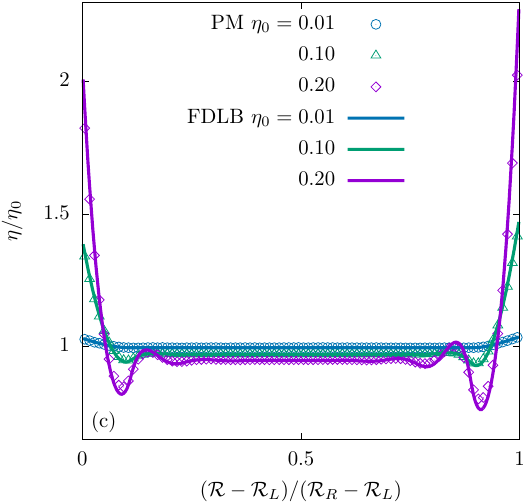}\\
\includegraphics[width=0.32\linewidth]{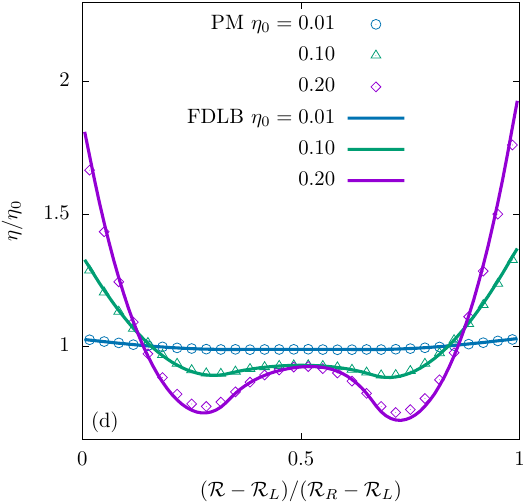}&
\includegraphics[width=0.32\linewidth]{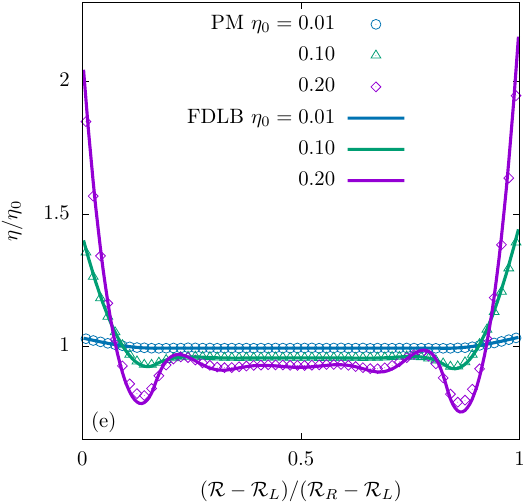}&
\includegraphics[width=0.32\linewidth]{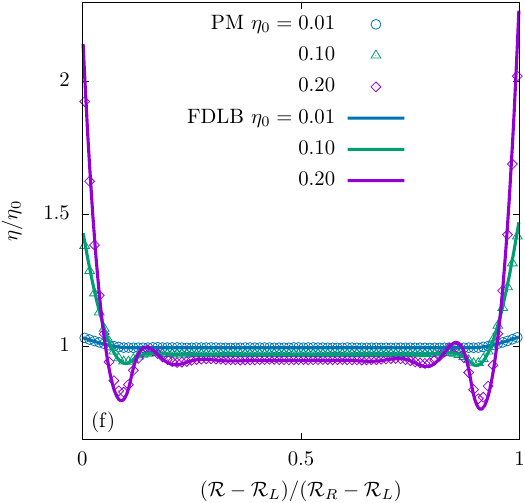}\\
\includegraphics[width=0.32\linewidth]{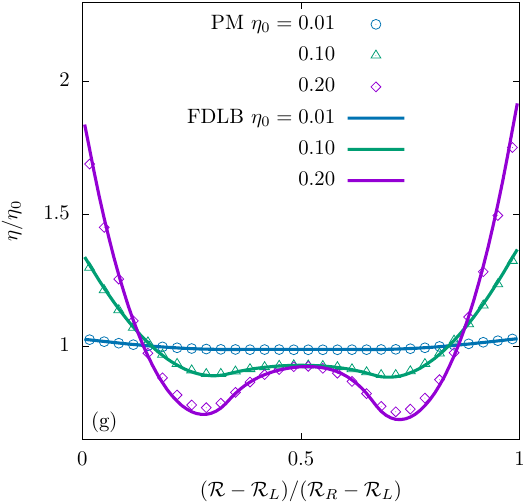}&
\includegraphics[width=0.32\linewidth]{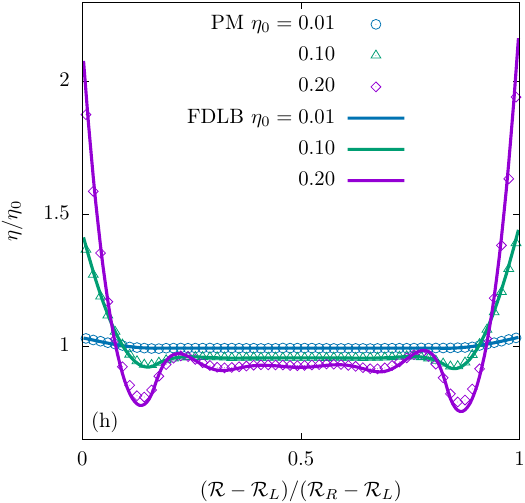}&
\includegraphics[width=0.32\linewidth]{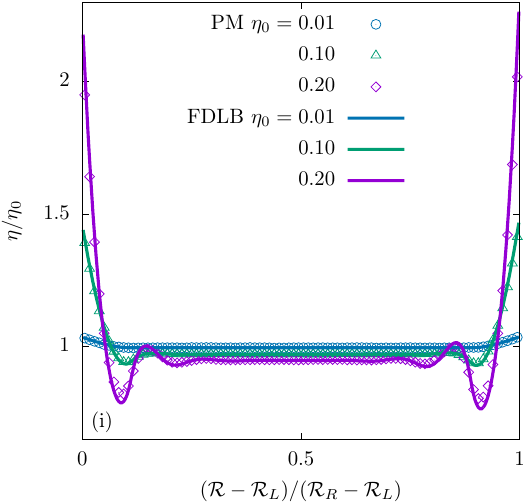}\\
\end{tabular}
\caption{Gas at rest: Normalized reduced density $\eta/\eta_0$ between coaxial cylinders and three values of the mean reduced density $\eta_0\in\{0.01,0.1,0.2\}$, three values of the inner cylinder radius $\mathcal{R}_L\in\{1,3,5\}$ (each row) and three values of the confinement ratio $C\in\{4,7,10\}$ (each column).\label{fig:stat_cyl_vareta}}
\end{figure*}

\begin{figure*}
\begin{tabular}{ccc}
4& 7& 10\\
\includegraphics[width=0.32\linewidth]{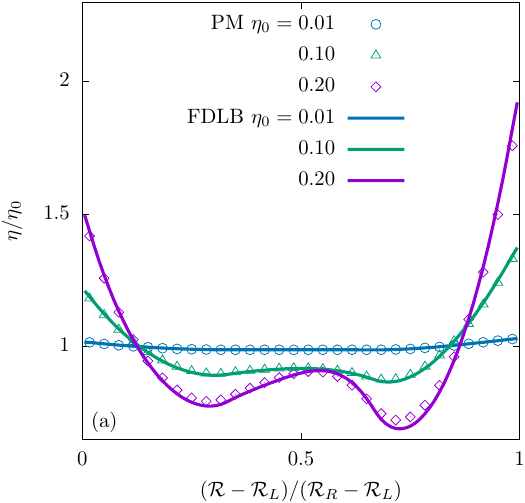}&
\includegraphics[width=0.32\linewidth]{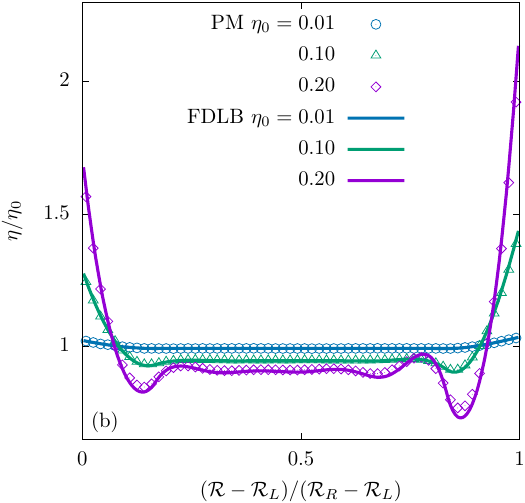}&
\includegraphics[width=0.32\linewidth]{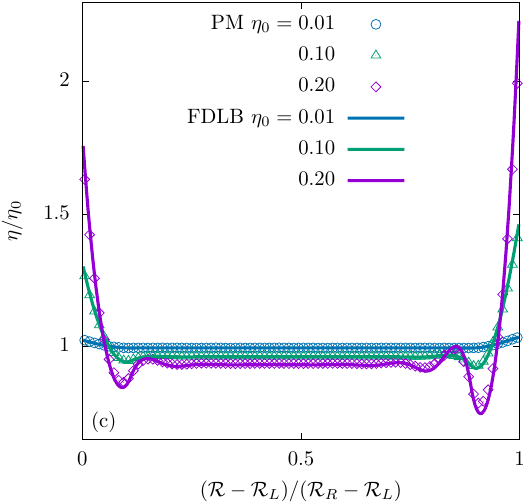}\\
\includegraphics[width=0.32\linewidth]{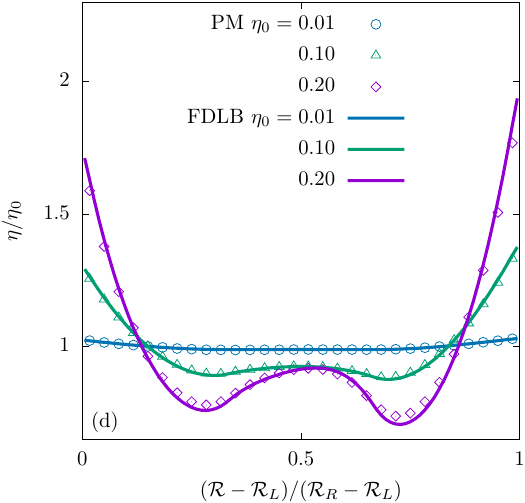}&
\includegraphics[width=0.32\linewidth]{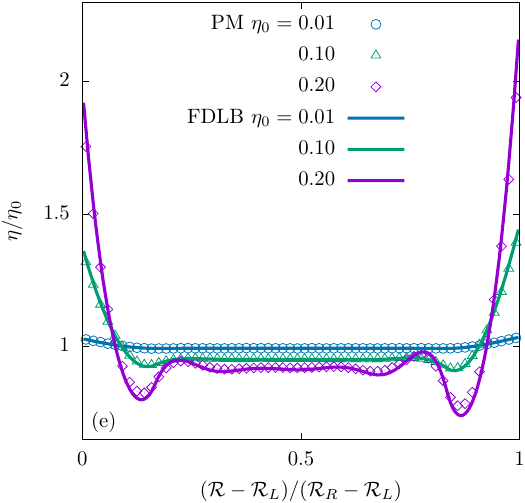}&
\includegraphics[width=0.32\linewidth]{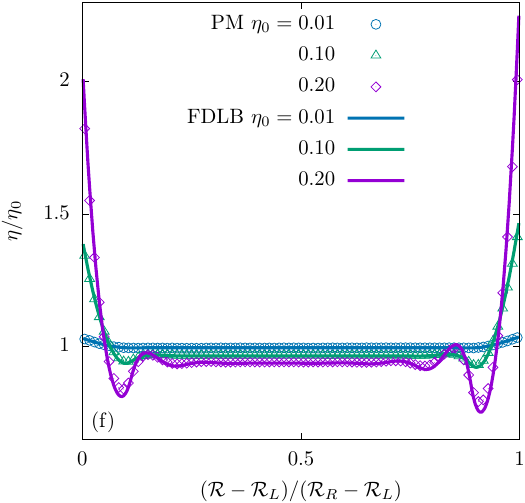}\\
\includegraphics[width=0.32\linewidth]{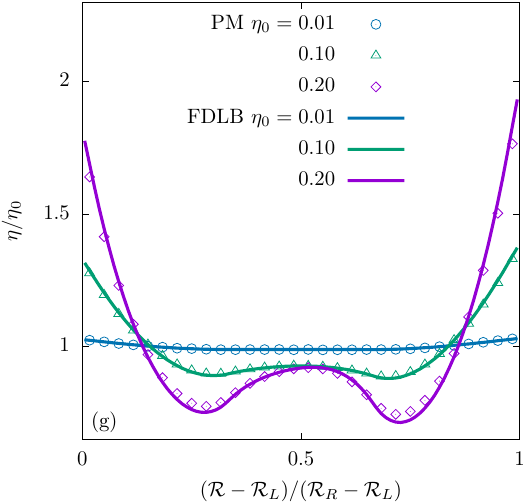}&
\includegraphics[width=0.32\linewidth]{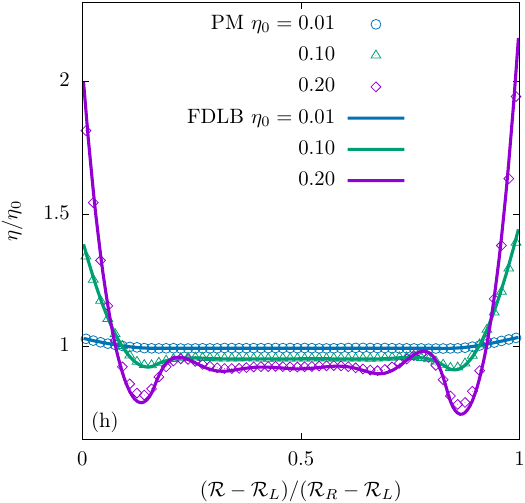}&
\includegraphics[width=0.32\linewidth]{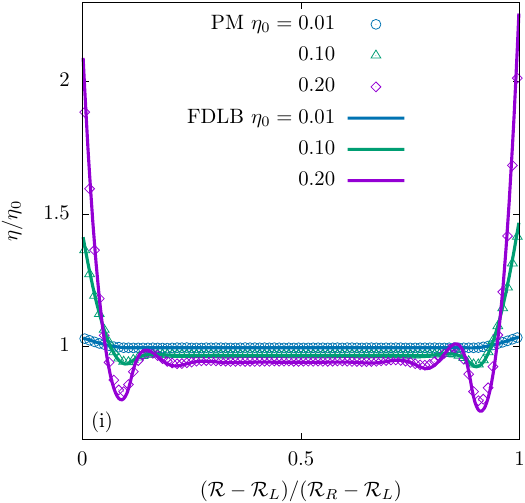}\\
\end{tabular}
\caption{Gas at rest: Normalized reduced density $\eta/\eta_0$ between concentric spheres and three values of the mean reduced density $\eta_0\in\{0.01,0.1,0.2\}$, three values of the inner cylinder radius $\mathcal{R}_L\in\{1,3,5\}$ (each row) and three values of the confinement ratio $C\in\{4,7,10\}$ (each column). \label{fig:stat_sph_vareta}}
\end{figure*}

\begin{figure*}
\begin{tabular}{ccc}
\includegraphics[width=0.32\linewidth]{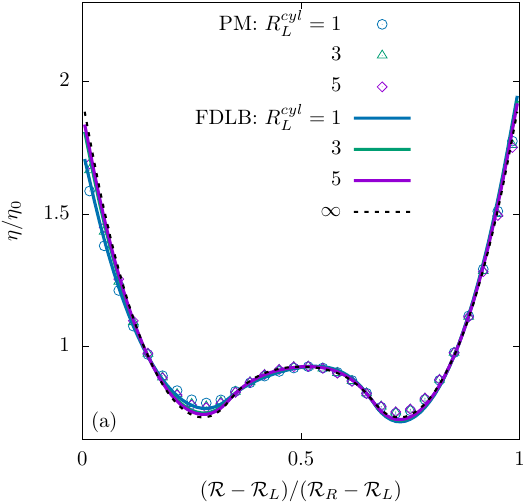}&
\includegraphics[width=0.32\linewidth]{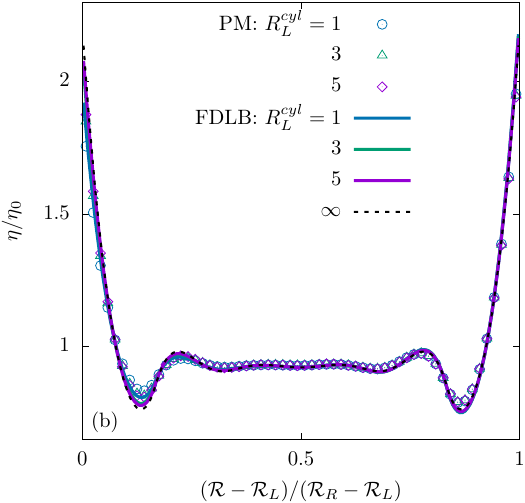}&
\includegraphics[width=0.32\linewidth]{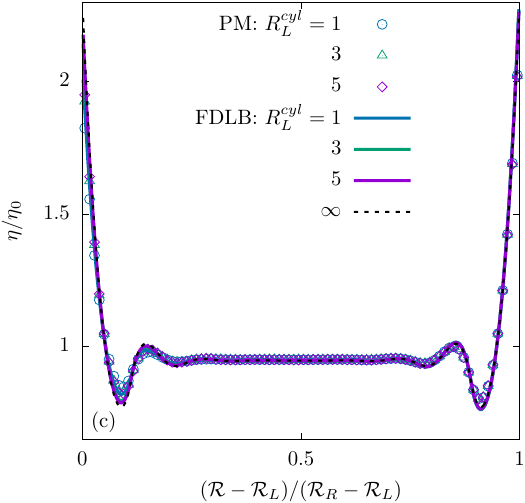}
\end{tabular}
\caption{Gas at rest: Normalized reduced density $\eta/\eta_0$ between coaxial cylinders and three values of the inner cylinder radius $\mathcal{R}_L$, the fixed mean reduced density $\eta_0=0.2$ and three values of the confinement ratio $C\in\{4,7,10\}$ (a-c). \label{fig:stat_cyl_varRl}}
\end{figure*}

\begin{figure*}
\begin{tabular}{ccc}
\includegraphics[width=0.32\linewidth]{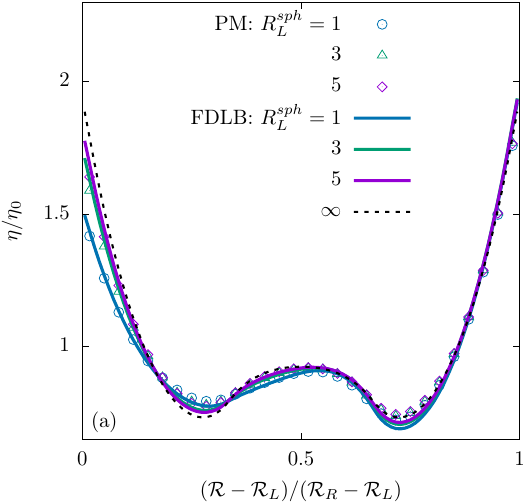}&
\includegraphics[width=0.32\linewidth]{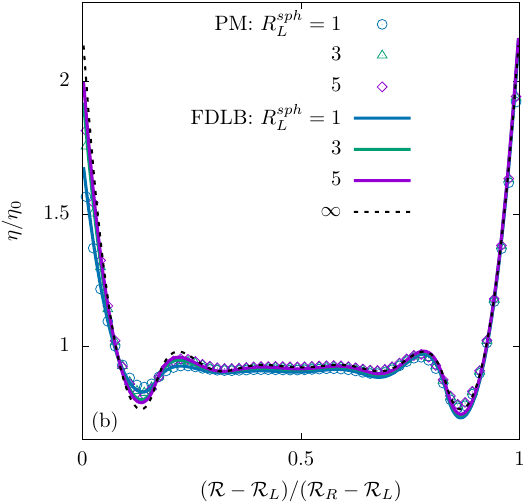}&
\includegraphics[width=0.32\linewidth]{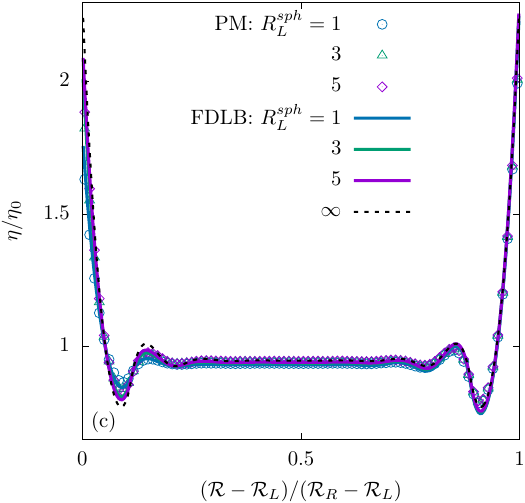}
\end{tabular}
\caption{Gas at rest: Normalized reduced density $\eta/\eta_0$ between concentric spheres, three values of the inner sphere radius $\mathcal{R}_L$, the fixed mean reduced density $\eta_0=0.2$ and three values of the confinement ratio $C\in\{4,7,10\}$ (a-c).\label{fig:stat_sph_varRl}}
\end{figure*}

\begin{figure}
\begin{tabular}{ccc}
\includegraphics[width=0.32\linewidth]{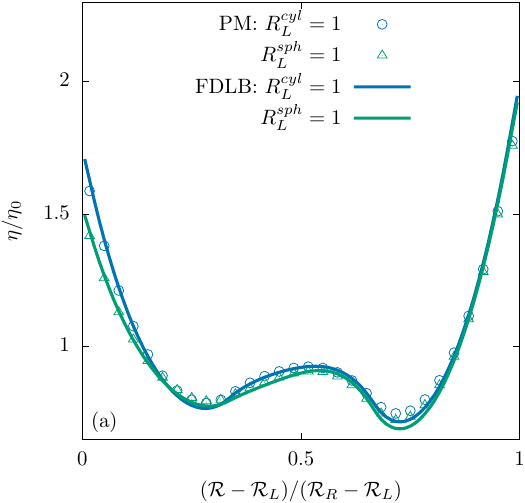}&
\includegraphics[width=0.32\linewidth]{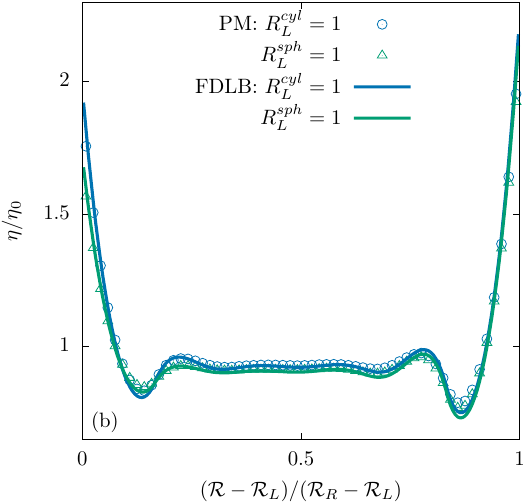}&
\includegraphics[width=0.32\linewidth]{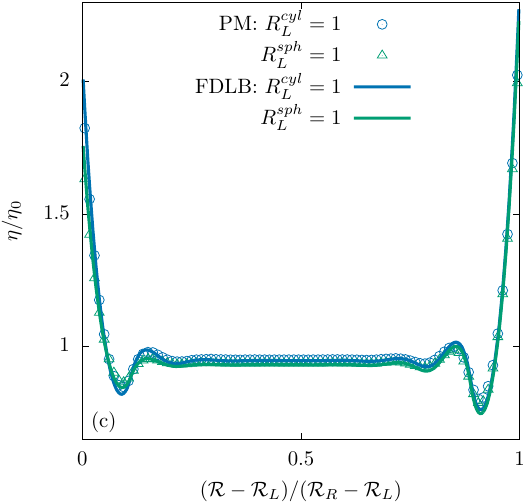}
\end{tabular}
\caption{Gas at rest: Normalized reduced density $\eta/\eta_0$ at $\mathcal{R}_L=1$ and variable geometry, for fixed mean reduced density $\eta_0=0.2$, fixed inner cylinder/sphere radius $\mathcal{R}_L=1$ and three values of the confinement ratio $C\in\{4,7,10\}$ (a-c).\label{fig:stat_varcoord}}
\end{figure}

In this Appendix, we compile the numerical results obtained for a hard-sphere gas at rest confined between two coaxial cylinders and two concentric spheres, kept at the temperature $T_w = 1$. We look at the reduced density profile in the channel and how it is impacted by various system setup variables. More specifically we choose three values of the initial reduced density $\eta_0\in\{0.01,0.1,0.2\}$, three values of the inner cylinder/sphere radii $\mathcal{R}_L\in\{1,3,5\}$ and three values of the channel width, i.e. the distance between cylinders/spheres, denoted using the confinement ratio $C\in\{4,7,10\}$, such that the outer cylinder/sphere radius is $\mathcal{R}_R=\mathcal{R}_L+C$.

Figs. \ref{fig:stat_cyl_vareta} and \ref{fig:stat_sph_vareta} present the normalized reduced density $\eta/\eta_0$ for a dense gas between coaxial cylinders and concentric spheres, compiled as varying confinement ratio $C\in\{4,7,10\}$ along each row and varying inner cylinder radius $\mathcal{R}_L\in\{1,3,5\}$ along each column, for three values of the initial reduced density $\eta_0\in\{0.01,0.1,0.2\}$. As was the case for planar wall \cite{BS24}, the stationary profile of the normalized reduced density $\eta/\eta_0$ is non-monotonic near the wall, a characteristic feature of dense gases, albeit its magnitude depends on the inner cylinder/sphere radius, i.e. the curvature. It is important to consider that when a fluid molecule is located at a distance less than a molecular diameter $\sigma$ from the wall, a portion of its surface remains protected from collisions since there is not sufficient space available for a second molecule to occupy that part of the spatial domain. As a result, the particle is pushed toward the wall. When dealing with a curved boundary this effect is either diminished or enhanced if a concave or convex boundary is involved. In our case, the more pronounced effect is on the inner cylinder/sphere where the available space is increased due to the shape of the boundary. As such the layering effect is inversely proportional to the curvature. Their intensity also diminishes as the dilute gas limit is approached $\eta_0\rightarrow 0$. These density variations emerge within a region approximately equivalent to the molecular diameter $\sigma$.

Furthermore, in order to compare to the planar wall limit we made plots with varying inner radius $\mathcal{R}_L\in\{1,3,5\}$, fixed confinement ratio $C$, and a mean reduced density $\eta_0=0.2$, for which the density variations are the largest. The results are displayed in  Figs. \ref{fig:stat_cyl_varRl} and \ref{fig:stat_sph_varRl}. One can observe that at a fairly small inner radius of the cylinder, the planar wall $\mathcal{R}_L\rightarrow\infty$ result is almost recovered, while for the sphere one needs to go a value larger than $\mathcal{R}_L>5\sigma$ in order to obtain the overlap.

Finally, we look at how the two geometries compare to each other for the same parameters. Fig. \ref{fig:stat_varcoord} present the results for $\mathcal{R}_L=1$, $\eta=0.2$ and three values of the confinement ratio $C\in\{4,7,10\}$. As expected, the layering at the inner wall is smaller in magnitude due to varying collision surfaces.

\bibliography{bibliography.bib}

\end{document}